\documentclass{pasj00}
\usepackage{color}

\author{
Kazuhiro {\sc Nakazawa}, \\
{\small {\it Department of Physics, The University of Tokyo,
                7-3-1 Hongo,  Bunkyo-ku,  Tokyo 113--0033}}\\
\centerline{\small {\it E-mail(KN) : nakazawa@amalthea.phys.s.u-tokyo.ac.jp}}\\
Craig L. {\sc Sarazin}, \\
{\small {\it Department of Astronomy, University of Virginia,
                P.O. Box 400325, Charlottesville, VA 22904-4325}}\\
Madoka {\sc Kawaharada},\\
{\small {\it Cosmic Radiation Laboratory, RIKEN,
                2-1 Hirosawa,  Wako,  Saitama 350--0198}}\\
Takao {\sc Kitaguchi}, Sho {\sc Okuyama}, Kazuo {\sc Makishima},\\
{\small {\it Department of Physics, The University of Tokyo,
                7-3-1 Hongo,  Bunkyo-ku,  Tokyo 113--0033}}\\
Naomi {\sc Kawano}, Yasushi {\sc Fukazawa}, \\
{\small {\it Department of Physical Science, Hiroshima University,
1-3-1 Kagamiyama, Higashi-hiroshima 739-8526}}\\
Susumu {\sc Inoue}, \\
{\small {\it 
Department of Physics, Kyoto University, Kyoto, Oiwake-cho,
Kitashirakawa, Sakyo-ku, Kyoto 606-8502}}\\
Motokazu {\sc Takizawa},\\
{\small {\it Department of Physics, Yamagata University,
1-4-12 Kojirakawa-machi, Yamagata 990-8560}}\\
Daniel R. {\sc Wik}, \\
{\small {\it Department of Astronomy, University of Virginia,
                P.O. Box 400325, Charlottesville, VA 22904-4325}}\\
Alexis {\sc Finoguenov}, \\
{\small {\it Max-Planck-Institut f\"ur Extraterrestrische Physik,
Giessenbachstra\ss e, 85748 Garching, Germany}}\\
{\small {\it also University of Maryland, Baltimore County, 1000
 Hilltop Circle,  Baltimore, MD 21250, USA}}\\
and\\
Tracy E. {\sc Clarke} \\
{\small {\it Naval Research Laboratory, 4555 Overlook Ave. SW, Code 7213, Washington D.C. 20375}}\\
{\small {\it also Interferometrics Inc., 13454 Sunrise Valley Drive, Suite 240, Herndon, VA 20171 }}
}

\begin{document}
\SetRunningHead{K. Nakazawa et al.}{Hard X-ray Properties of Abell~3667 as Observed with {\it Suzaku}}
\Received{} %{2002 September 4}
\Accepted{}     %{2003 March 1}
\title{Hard X-ray Properties of the Merging Cluster Abell~3667 as Observed with {\it Suzaku}
}

\maketitle

\begin{abstract}

Wide-band {\it Suzaku} data on the merging cluster Abell 3667 
were examined for 
hard X-ray emission in excess to the known thermal component.
{\it Suzaku} detected X-ray signals in the wide energy band from 0.5 to 40 keV.
The hard X-ray ($> 10$ keV) flux observed by the HXD around the cluster center
cannot be explained by a simple extension of the thermal emission with average temperature of $\sim 7$~keV.
The emission is most likely an emission from
a very hot ($kT > 13.2$ keV) %10.6 linear --> 12.9 quadrature
thermal component around the cluster center,
produced via a strong heating process in the merger.
In the north-west radio relic, no signature of non-thermal emission was observed.
Using the HXD, the overall upper-limit flux within a $34'\times34'$ field-of-view around the relic
is derived to be 5.3$\times10^{-12}$ %3.2--7.0 lin --> 1.7--5.5
erg s$^{-1}$ cm$^{-2}$ in the 10--40 keV band,
after subtracting the ICM contribution estimated using the XIS or the {\it XMM-Newton} spectra.
Directly on the relic region, the upper limit is further tightened by the XIS data to 
be less than $7.3\times10^{-13}$ erg s$^{-1}$ cm$^{-2}$, %8.1 lin --> 6.8 sq
when converted into the 10--40 keV band.
The latter value suggest that the average magnetic field within the relic
is higher than $1.6~\mu$G. %1.5 lin -> 1.6 sq
The non-thermal pressure 
due to magnetic fields and relativistic electrons may be 
as large as $\sim 20$\% of the thermal pressure in the region.
\end{abstract}

%%% 1. Introduction %%%%%%%%%%%%%%%%%%%%%%%%%%%%%%%%%%%
\section{Introduction}
%%%%%%%%%%%%%%%%%%%%%%%%%%%%%%%%%%%%%%%%%%%%%%%%%%%%%%%

Enigmatic extended radio sources in rich galaxy clusters
have been known for over 30 years (Willson 1970; for recent review, see Feretti 2005).
Sources located at the cluster center are referred to as ``radio halos'',
while those on the cluster periphery are called ``relics''.
Steep spectra and spectral cutoffs at a few GHz detected in several halos/relics 
indicate that relativistic electrons with typically GeV energy are 
undergoing cooling due to synchrotron and inverse Compton (IC) emission.
In every case, the radio halos/relics are found in irregular clusters which are
apparently undergoing mergers. This suggests that the radio emitting electrons are
accelerated by shocks or turbulence generated by energetic cluster merging.

The same relativistic electrons scatter
Cosmic Microwave Background (CMB) photons up to the hard X-ray band (IC emission).
By comparing the radio and the hard X-ray emissions, 
not only the electron energy density but also the intra-cluster magnetic
field intensity can be estimated (e.g. Sarazin 1988).
Detecting the IC emission is difficult because of
the strong ICM component dominant below $\sim 20$ keV
and the low expected flux of the emission in hard X-rays above this energy.
Currently, there are only a few reports claiming its detection,
mostly from the {\it Beppo-SAX} PDS in the energy band around 40 keV or higher 
(e.g. Fusco-Femiano et al.~1999; Nevalainen et al.~2004),
though a few of the detections are still controversial 
(Rossetti \& Molendi 2004; Fusco-Femiano et al.~2007).
There are reports from RXTE (e.g. Petrosian et al.~2006; Rephaeli et al. 2006) as well.
Another case includes the {\it ASCA} GIS results around 4 keV on galaxy groups
(Fukazawa et al.~2001; Nakazawa et al.~2006).
These results would require a magnetic field as low as 0.1 $\mu$G (e.g. Fusco-Femiano et al.~1999),
which at the first glance is inconsistent with the 
radio rotation measure estimation of $2\sim10$ $\mu$G 
(e.g. Clarke et al.~2001; Carilli \& Taylor 2002).
Further observations by independent instruments of IC and other 
hard X-ray emission processes (e.g. Inoue et al.~2005), is thus desirable.

Significant heating of the ICM itself must also be taking place
in merging clusters (e.g. Takizawa 2000; Ricker \& Sarazin 2001).
X-ray hardness ratio images sometimes show 
temperatures exceeding 10 keV (e.g. Watanabe et al.~1999; Briel et al.~2004).
Since the emission can be multi-phase within the line of sight,
the actual highest temperature can be much higher,
though it is not easy to identify using the contemporary imaging X-ray detectors
working below $< 12$~keV.

In many ways, Abell 3667 is the ideal cluster to study mergers, radio relics and hard X-ray emission.
It is a very bright X-ray cluster at a low redshift of $z = 0.0556$ (Struble \& Rood 1999). 
The {\it ROSAT} and {\it ASCA} observations 
showed that it is a spectacular merger with shock heated gas (Markevitch et al.~1999).
{\it Chandra} and {\it XMM-Newton} observations have 
shown much evidence for an ongoing merger,
such as a cold front (Vikhlinin et al.~2001a,b) and
highly inhomogeneous temperature structure with temperature ranging from 4 to $> 10$ keV
(Mazzotta et al.~2002; Briel et al.~2004).
The optical galaxy distribution shows elongation towards 
the north-west south-east axis (Sodre et al.~1992; Johnston-Hollitt et al.~2008),
supporting the binary merger scenario as well.

The cluster is famous for its pair of radio relics (e.g. Roettgering et al. 1997).
The relic to the north-west is the brightest and largest among the diffuse radio sources associated 
with cluster of galaxies, 
having a flux of 3.7 Jy (Johnston-Hollit 2004) and a width of $\sim 20'$ at 1.4 GHz.
It is located about $30'$ or 2 Mpc from the cluster X-ray centroid.
The radio signal is detected from 85 MHz to 2.3 GHz,
and its average photon index is $\Gamma = 2.1$ (Roettgering et al. 1997).
Here the photon index $\Gamma$ is defined as $N_{\rm photon}(E) = N_0 \times E^{-\Gamma}$.
The radio relics have very sharp outer edges and the radio spectra steepen with distance from the edge.
Thus, the radio relics are considered to reflect the position of the particle acceleration in a merger shock.
The cluster is also a good candidate for detection of IC emission.
The {\it Beppo-SAX} PDS provided a rather marginal evidence of hard excess emission from the cluster
(Fusco-Femiano et al. 2001; Nevalainen et al. 2004). 
However, the large field of view (FOV) of the PDS makes the data strongly affected by contamination from 
thermal emission from the entire cluster, mainly from its central region, and possible AGNs.

% << reference update at 080708

The hard X-ray detector (HXD; Takahashi et al. 2006; Kokubun et al. 2006) onboard 
{\it Suzaku} (Mitsuda et al. 2006)
is characterized by its low detector background in the 10--40 keV band
and its narrow FOV of $34'$ full width at half maximum (FWHM).
Using the HXD, we can spatially distinguish hard X-ray components 
from the cluster center and the north-west relic, with the highest sensitivity in the 10--40 keV band.
In addition, the X-ray CCD cameras (XISs; Koyama et al. 2006) onboard {\it Suzaku}
are characterized by their large effective areas and low and stable detector backgrounds,
which make them very powerful devices to observe extended hard emission.

In May 2006, we observed Abell 3667 cluster using {\it Suzaku} with three pointings
running from the north-west relic to the center, each separated by $\sim 17'$.
%Total exposure was 105 ks for the XIS. 
Here we report the results obtained by these observations.
Observation logs and data reduction are described in the next section.
Section \ref{chap:spec} is devoted to the spectral and imaging analysis,
followed by the discussion in section \ref{chap:discussion} 
and conclusions in section \ref{chap:conclusion}.
In this paper, we utilized canonical cosmology value of 
$H_{0} = 70$ Mpc km$^{-1}$s. 
At the redshift of $z = 0.0556$, $1'$ corresponds to 62 kpc. 
For a mean cluster temperature 
of 7.2 keV, the virial radius should be $r_{200} \sim 2.6$ Mpc $= 42'$ (e.g. Neumann \& Arnaud 1999).
Unless otherwise noted, all errors are at the 90\% confidence limit. 

%%% 2. Observation %%%%%%%%%%%%%%%%%%%%%%%%%%%%%%%%%%%%
\section{Observation and Data Reduction}
\label{chap:obs_and_data_reduction}
%%%%%%%%%%%%%%%%%%%%%%%%%%%%%%%%%%%%%%%%%%%%%%%%%%%%%%%
%--------- 2.1 ------------------------
\subsection{Observation logs}
%--------- 2.1 ------------------------

%%%######################### table 1 #######################################

%             start               end
%center 2006-05-06T17:40:17 2006-05-07T09:13:24
%17off  2006-05-06T07:03:14 2006-05-06T17:39:24
%relic  2006-05-03T17:47:01 2006-05-06T07:02:25

\begin{table}[htbp]
\caption{Log of the three Abell~3667 observations.}
\label{table:obs-date}
  \begin{center}
    \begin{tabular}{lccc}
\hline %-----------------------------------------------------------------
\hline %-----------------------------------------------------------------
Obs.  & Start (UT)  &  End (UT) &  Exp. (ks) \footnotemark[$*$]  \\
\hline %-----------------------------------------------------------------
NW\_Relic & 05/03 17:47 & 05/06 07:02 & 81.2/45.6\\
NW\_17off & 05/06 07:03 & 05/06 17:39 & 16.7/11.4 \\
Center  & 05/06 17:40 & 05/07 09:13 & 20.5/7.2  \\
\hline %-----------------------------------------------------------------
\end{tabular}
\end{center}
  \footnotesize
\footnotemark[$*$] The XIS/HXD exposure time after the screening as presented in the text.\\
  \normalsize
\end{table}
%%%########################################################################

The three Abell~3667 pointing observations of {\it Suzaku} were carried out 
successively in the beginnng of May 2006. 
As already noted, individual pointing positions are offset by $\sim 17'$,
so that the first observation covers the north-west relic,
and the last one on the cluster center, with continuous XIS coverage.
Hereafter, we call these observations, the NWR, the 17$'$ offset and the center pointing, respectively.
Details of the observation time are summarized in table~\ref{table:obs-date}. 
The FOVs of individual pointings compared with 
the X-ray sky image of PSPC are also presented in appendix \ref{apdx:point_src}.

The XIS was operated in the normal timing and full window mode.
All the 64 PINs of the HXD were operated in the nominal bias voltage of 500 V. 
All the data was processed with the {\it Suzaku} pipeline processing of version 2.0.6.13.

%--------- 2.2 ------------------------
\subsection{Data reduction of the XIS}
%--------- 2.2 ------------------------

\subsubsection{Screening and background estimation}
\label{chap:xis_screen_bgd}

The XIS data were processed via default screening criteria.
Events with a GRADE of 0, 2, 3, 4, 6 and STATUS $<$ 1024 were extracted.
Data screening criteria with HK files are as follows: 
time after SAA (T\_SAA\_HXD) $> 436$~s, the new cut-off rigidity (COR2) $> 6.0$ GV,
elevation from the earth rim $>5$\degree ~ and that from the sun-lit earth rim $> 20$\degree.
Non X-ray background (NXB) spectra of the XIS were created by sorting
night earth data by COR2, and weighted-averaging
them using the ftool ``xisnxbgen'' (Tawa et al.~2008).
The estimation uncertainty in the NXB level is 
handled as a systematic error in the following analysis.
Tawa et al.~(2008) studied the error on a typical observation lasting for a few days.
At the 90\% confidence level, it is 6.0\% and 12.5\% for the sum of three  %new 080607
XIS-FI (XIS-0,2,3) data and the BI (XIS-1) data, respectively.

Another major background is the Cosmic X-ray background (CXB).
Details of the CXB estimation is shown in appendix \ref{apdx:xis_cxb}.
Because the whole XIS region in all the three observations could 
be filled with non-thermal or thermal diffuse emission, 
we used the Lockman Hole observation ({\it Suzaku} observation ID, 101002010) as the CXB template.
In addition, the northern 1/3 of the NWR pointing data 
was utilized for foreground Galactic soft thermal component modeling.
The CXB fluctuation is estimated to be 11\% over the XIS full FOV.
Hereafter, we treat the background-subtracted (both the NXB and CXB) 
spectra as signals from Abell~3667.

\subsubsection{The XIS image}

\begin{figure}
  \begin{center}
    \FigureFile(7.5cm,7.5cm){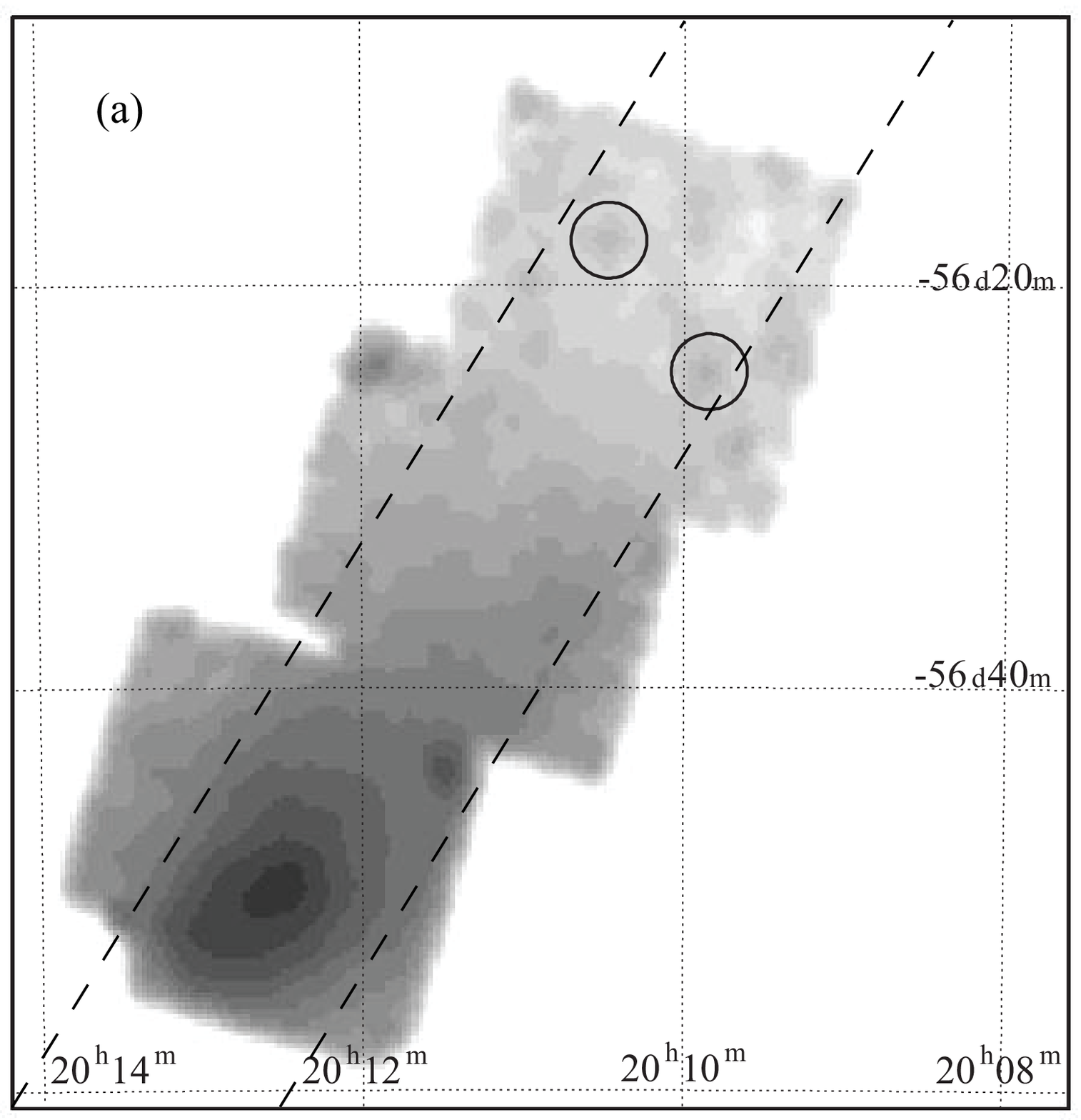}       %%% 
    \FigureFile(7.5cm,7.5cm){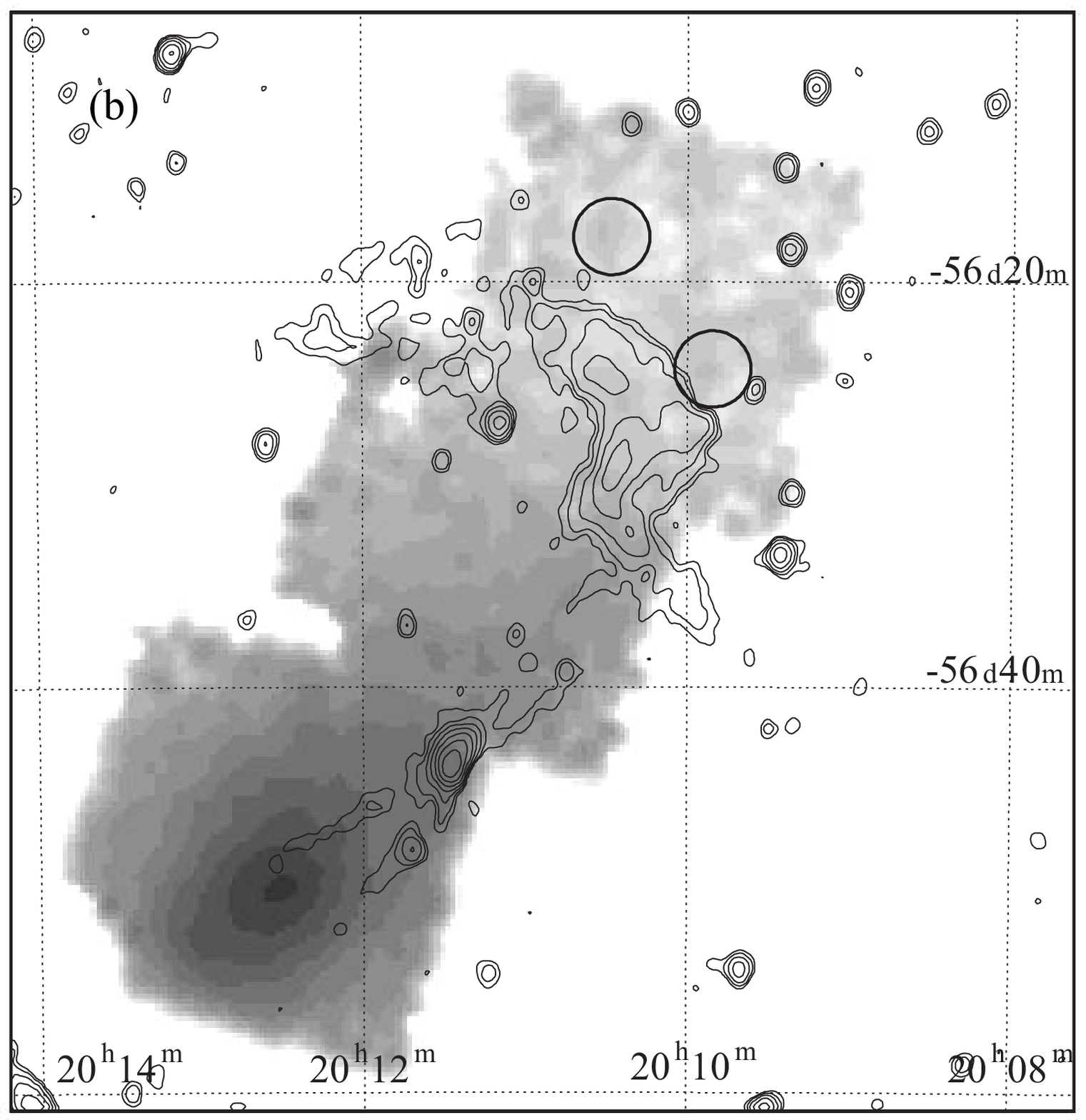}       %%% 
  \end{center}
\caption{Mosaic XIS image of the three pointing observations, in (a) the soft 1-4 keV band
and (b) the hard 4-8 keV band. See text for details.
The images are smoothed by a Gaussian kernel with $\sigma = 0.5$ arcmin. 
Gray-scale contours are logarithmically scaled by a factor of 1.5.
In the hard band image, the 843 MHz radio image (by SUMSS) is overlaid as a thin contour.
Two point source candidates are shown as open circles in the NWR region,
and the projection analysis band is shown as a dashed lines in panel (b);
both are discussed later in section \ref{chap:nwr_xis}.
}
\label{fig:xis_mosaic}
\end{figure}

In figure~\ref{fig:xis_mosaic}, we show 1--4 keV and 
4--8 keV XIS-FI mosaic images of the three pointings.
After co-adding the images from XIS-0, 2 and 3,
they were corrected for the overlapping exposure,
and for vignetting effect of the mirror optics, after subtracting the NXB.
Thanks to the wide energy range and low background of the XIS, 
we can see signals even in the NWR region 
which is $30' ~( 2$ Mpc) away from the X-ray centroid.
On the other hand, no apparent emission associated with the 
radio relic is seen in neither band image.
Detailed analysis using these images is performed in section \ref{chap:nwr_xis}.

%%#################### figure 1 #########################################
%%#######################################################################

%--------- 2.3 ------------------------
\subsection{\it Data reduction of the HXD }
%--------- 2.3 ------------------------

\subsubsection{Screening and background estimation}
\label{chap:hxd_screen_bgd}

We processed the uncleaned event files of the HXD with ``hxdgtigen'', 
and remaining events were further screened with HK files as follows;
time after SAA (T\_SAA\_HXD) $> 500$ s, the COR2 $> 8.0$ GV
and elevation from the earth rim $>5$\degree.
As the NXB of the PIN, we utilized public NXB model which is provided
by the HXD team in the {\it Suzaku} website\footnote{http://www.astro.isas.jaxa.jp/suzaku/analysis/hxd/pinnxb/}.
The version of the model is ``METHOD=LCFITDT'', or ``tuned''.
Observations of Abell~3667 was carried out in a period in which the lower threshold of 
the scintillator was set low, and a data transfer overflow between the 
HXD's analog and digital electronics sometimes occurs. 
This phenomena should not take place in ordinary observations,
and hence corresponding time regions are discarded in the default screening criteria.
In fact, the effect of this phenomena in principle is recoverable through 
the dead-time correction procedure and thus the data should be useful.
Since the HXD data in this paper, however, are dominated by systematic errors and not statistics,
and most of the systematic error analysis is performed
in periods excluding these phenomena, we stick to the original screening criteria.
Note that none of the results are changed significantly if we include the data obtained in periods with
the data transfer overflow.

In the HXD-PIN NXB documents 
(Fukazawa et al.~2008; Mizuno et al.~2008\footnote{{\it Suzaku}-memo, JX-ISAS-SUZAKU-MEMO-2008-03.}),
the reproducibility of blank-sky observations separated into 10 ks
exposures gave a distribution of 5.7\% at the 90\% confience level, including the statistical error of
typically 3.3\% or larger and the CXB fluctuation of 1.3\%, as described later.
This gives the NXB systematic uncertainty of 4.5\% at the 90\% confidence. 
In the following analysis, we utilized this value as the systematic error of the NXB.
Because the selection criteria of the blank-sky observations are simple and based only on the XIS source flux,
some of them can be contaminated by off-center sources. In addition, 
10 ks is a bit shorter than the exposure of our observations.
Thus, the actual systematic error is probably smaller than this value.

Detail of the CXB model used for the PIN is described in appendix \ref{apdx:hxd_cxb}.
We defined the photon flux model as
$N(E) = 8.69\times10^{-4} \times (E)^{-1.29} \times {\rm exp}(-E/40.0)$ 
in ${\rm photons~ cm^{-2}~s^{-1}~keV^{-1}~FOV^{-1} }$.
Here, $E$ is the photon energy in keV,
and the model is normalized to the HXD-PIN opening angle (or FOV) of 
$\Omega_e^{\rm HXD} = 0.32$ deg$^2$, to be combined with the HXD nominal response.
The 13.2\% systematic difference of the XIS and PIN 
(Ishida et al. 2007)\footnote{{\it Suzaku}-memo, JX-ISAS-SUZAKU-MEMO-2007-11.} is also accounted
in the following fitting process.
The level of the CXB fluctuation is calculated to be 18\%,
which corresponds to 1.3\% of the CXB+NXB in the 10-40 keV band.

We also checked whether the PIN data suffered 
from thermal noise in the lower energy band by
comparing the actual earth occultation data with the corresponding model NXB spectra.
Since the detector temperature was relatively high at $-12$ C$^{\circ}$ in these observations,
the thermal noise is seen as a steep rise below 13 keV.
In the higher energy band, the two spectra are consistent within statistics.

\subsubsection{Signals of the three pointings}
\label{chap:hxd_3c heck}

\begin{figure}
  \begin{center}
\FigureFile(7.5cm,5cm){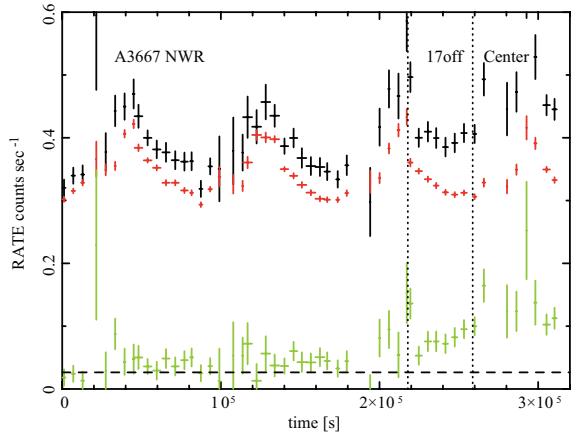}       %%% 
  \end{center}
  \caption{The HXD light curves in the 13-40 keV range. Black crosses stands for the data, red for the background model and green for the residual signals. 
Time regions of the three observations are separated by vertical dotted lines.
The horizontal dashed line indicates the expected CXB rate.
}
\label{fig:hxd_lc}
\end{figure}

In figure~\ref{fig:hxd_lc}, the 13--40 keV band light curves 
of PIN for a total of 4 days of observations are presented.
The PIN background varies by $\sim 25$\%,
mainly correlated with COR, and time elapsed after a passage of SAA (T\_SAA\_HXD).
The PIN residual signal clearly exceeds the CXB level for all three observations.
Note that the flux increases towards the cluster center.
Small variations seen in the residual signal are due to uncertainties in 
the NXB modeling in this rather short time bin (90 minutes),
and the NXB-subtracted count rate is consistent with being constant within each pointing.

%%#################### figure 1 #########################################
%%#######################################################################

\begin{figure}
  \begin{center}
     \FigureFile(6.5cm,9cm){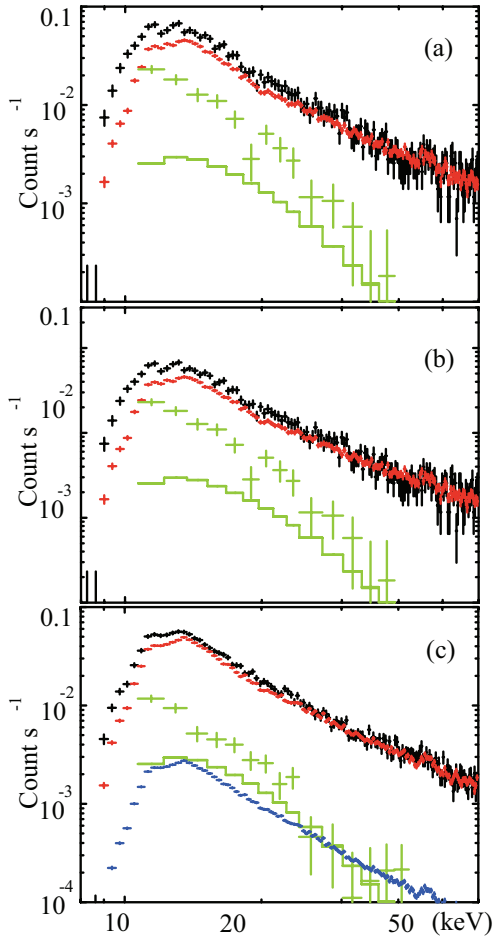}
  \end{center}
  \caption{PIN spectra of the three pointings, from (a) center,
(b) 17$'$ off and (c) NWR. Black crosses stand for the data, 
red for the NXB model and green for the residual signals (data $-$ NXB). 
The expected CXB spectra are also plotted as green histograms.
Typical systematic errors due to the CXB fluctuations and the NXB reproducibility 
are shown in the NWR data in blue.
}
\label{fig:hxd_raw_spec}
\end{figure}

In Figure~\ref{fig:hxd_raw_spec}, we show raw PIN spectra from 
the 3 observations, together with the NXB model spectra.
Difference between the two spectra are also presented,
and compared to the CXB model.
In all cases, the NXB-subtracted data significantly exceeds the CXB.
The signal is strongest in the center pointing, and weakest in the NWR pointing.
The 13--40 keV PIN signal rates are  
$(8.9\pm1.3$,
$5.4\pm1.0$ and $1.4\pm0.5) \times 10^{-2}$~cts~s$^{-1}$
in the center, the $17'$ offset, and the NWR data, respectively.
The systematic error is $1.5 \times 10^{-2}$~cts~s$^{-1}$
in all pointings, 
when evaluated as a quadrature sum of those of the CXB and the NXB.
Thus, we can claim the signal detection from Abell 3667 up to 30 keV
both in the center and 17$'$ offset observations.

%%#################### figure 1 #########################################

We also estimated the point-source contribution to the HXD-PIN
using the {\it ROSAT}-PSPC and the {\it ASCA}-GIS data. 
In short, we confirmed that the contribution from these sources is 
less than 10\% and hence almost negligible.
Details are shown in appendix \ref{apdx:point_src}.

%--------- 2.3 ------------------------
\subsection{Preparation of arf files of the XIS and the PIN}
%--------- 2.3  ------------------------
\label{chap:arfgen}

When trying to quantify the brightness and spectral shape of the PIN signals,
we need to know the angular transmission 
(or effective area) of PIN to the corresponding hard X-ray emission.
This in turn requires knowledge of the location and shape of the source, which is unavailable
in the PIN energy range.
In evaluating the ICM contribution to the PIN band, we therefore assumed it to have the same
spatial distribution as the ICM signals detected with the PSPC.
 Here, the ICM was assumed to be isothermal,
which is not completely true for this cluster.
However, this assumption is valid as a starting point,
since the major part of the ICM emission contributing to the PIN band originates 
from within $20'$ of the cluster center,
within which region the average spectra is well approximated by a $\sim 7$ keV ICM emission
(see sections \ref{chap:center_thermal_xis_pin} and \ref{chap:offset_chk}).
In this case, inhomogeniety in the ICM spectra is observable as a small shift in
spectra of individual pointings.

The PSPC image was background subtracted and
corrected for the vignetting effects following the method described in Snowden et al. (1994). 
Since the ICM emission in the image is not observable beyond $40'$ from the cluster center,
we discarded the data in this region.
After normalizing the image to have a total area of 1.0 and binning
it with a $2'\times2'$ mesh, we simply calculated the PIN arf file by convolving 
the point source arf of the specified mesh multiplied by its relative intensity.
Note that the energy response has in principle negligible
dependence on the position within the PIN-FOV.
The cross-normalization factor of $1.132$ between the PIN and the XIS
is incorporated, as well.

Regions beyond $26'$ from the cluster center,
i.e. those as far as the XIS FOV of the NWR pointing,
contribute only 1.3\%, 4.9\% and 14.4\% to the normalization of 
arf files of the center, the $17'$ offset and the NWR pointings, respectively.
Thus, 86\% of the ICM photon detected in the NWR PIN data
is expected to come from the inner region of the cluster.
Interestingly, the normalization relative to the center pointing
thus calculated are 0.61 and 0.14 in the 17$'$ offset and the NWR pointings, respectively.
This value matches the actually observed ratios,
$0.60\pm0.25$ and $0.11\pm0.18$, respectively.
As a reference, those of a point source located at the cluster X-ray centroid
are 0.42 and 0.05, respectively.
These results suggests that the ICM emission around 
the cluster central region must be carefully handled when analyzing 
the NWR PIN data.

In a similar way, we synthesized the XIS arf files with the ftool ``xissimarfgen'',
using the PSPC image as an input.
Note that the input image is automatically normalized to 1.0 in the software.
Thus, the spectral fitting of both the PIN and the XIS is normalized to 
the whole cluster, and not the FOV of each detector of individual observations.
Since the NWR region is around the detection limit of
both the PSPC and the XIS, the arf files for this region
could be inaccurate, and they require special care in the following analysis.
This effect is negligible in the center and 17$'$ offset pointings because
of the high X-ray surface brightness.

To verify the accuracy of the generated arf files,
we compared the XIS results to those of the PSPC.
We derived the 0.5--2.0 keV ICM flux by integrating the
PSPC surface brightness profiles given by Mohr et al.~(1999)
within $40'$ from the X-ray centroid.
The value matched within 1\% to the flux obtained by fitting the center XIS FI spectra using our arf,
which is also normalized to the PSPC image integrated over $40'$.
Here, a $kT = 7$ keV single temperature thermal emission was assumed.
Although the analysis in this paper is not sensitive to the {\it Suzaku}-PSPC
cross calibration, this result supports our arf generation procedure.
On the other hand, the XIS and PIN cross-calibration,
to which our analysis is sensitive, is verified within $\sim 3$\% accuracy 
using the Crab observations, including the mapping within the PIN-FOV.
Because the PIN-FOV is relatively simple in its shape, no strong systematic error is expected.
Using the similar approach, the PIN spectra 
of the Abell 1060, Centaurus (Kitaguchi et al. 2008) and Coma clusters (Wik et al. 2008)
were reproduced successfully by extrapolating the XIS or {\it XMM} spectra.
From these results, we conservatively assume a 10\% uncertainty in the 
{\it relative} normalization of the XIS and the PIN arf files.

%--------- 3 ------------------------
\section{Wide-band spectral analysis}
\label{chap:spec}
%--------- 3 ------------------------

\subsection{Spectral properties of the central pointing}
\label{chap:center}

With the preparation made in section \ref{chap:obs_and_data_reduction},
we proceed to the spectral analysis of the XIS and the PIN.
The central pointing afforded the signal detection with the highest significance 
using both instruments. Their arf files are also the most reliable since 
the brightest central emission is well detected by the PSPC.
Thus, we started the spectral fitting from the central pointing.
We first evaluated the XIS and the PIN spectra individually, 
and then performed the combined fitting.

Spectra from the three XIS-FI chips are summed up to form a single XIS-FI spectrum.
Since the XIS-BI chip has a significantly different detector response, its spectra is separately handled.
The arf files prepared in section \ref{chap:arfgen} are used 
for the XIS-FI, XIS-BI and the PIN data.
Thus, the normalization of individual spectra should be the same.
Practically, we fixed the relative normalization between the XIS-FI and the PIN data,
and the 10\% systematic uncertainty was handled separately.
The normalization between the XIS-FI and BI data was set free,
to adjust for residual mutual calibration uncertainties of about 5\% or less.
Here, the XIS-FI, the XIS-BI and the PIN spectra 
are utilized in 0.6--10.0 keV, 0.6--8.0 keV, and 13--40 keV, respectively.
Since the data have very high statistics, 
the energy range of 1.76--1.86 keV in the XIS spectra were ignored to avoid
uncalibrated structures caused by the Si edge.
The XIS-FI and BI spectra are grouped so that each energy bin has
at least 200 counts. 
The PIN spectra are grouped into fixed 12 bins, each with sufficient counts.

%\subsubsection{The spectra of the XIS and the PIN}
\subsubsection{The XIS}
\label{chap:center_thermal_xis_pin}

%%#################### figure 1 #########################################
\begin{figure}
\begin{center}
    \FigureFile(6cm,8cm){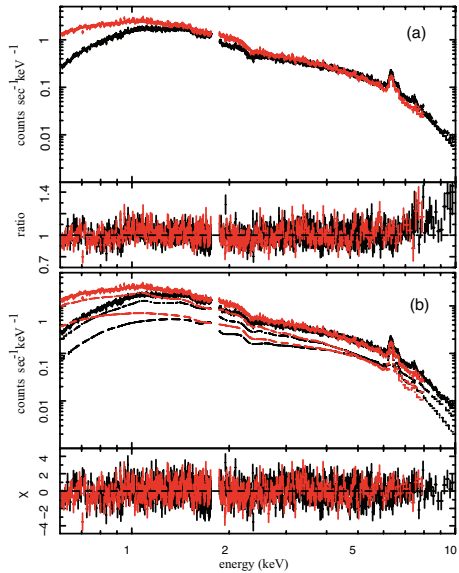} 
\end{center}
\caption{The XIS-FI (black) and XIS-BI (red) center spectra, fitted with (a) $1kT$ and (b) $2kT$
thermal emission model. In the panel (b), 
two thermal components are independently presented as dashed histograms.}
\label{fig:center_xis_spec1}
\end{figure}
%%#######################################################################

%%%######################### table #######################################
\begin{table}
  \caption{Best fit parameters of the $1kT$ and $2kT$ models fitted to the center XIS spectra.}
  \label{table:cenfit_xis}
  \begin{center}
    \begin{tabular}{lll}
\hline %-----------------------------------------------------------------
\hline %-----------------------------------------------------------------
 & $1kT$ & $2kT$ \\
 \hline %-----------------------------------------------------------------
$kT_{\rm low}$\footnotemark[$*$] & $6.8$& $4.71_{-0.34}^{+0.27}{~}_{-0.06}^{+0.07}$\\
$kT_{\rm hi}$\footnotemark[$\dagger$] & --& $22.3_{-5.0}^{+7.5}{~}_{-1.4}^{+1.1}$\\
$Z$\footnotemark[$\ddagger$] & $0.32$ & $0.38_{-0.03}^{+0.03}{~}_{-0.00}^{+0.01}$\\
$N_{\rm low}$\footnotemark[$\S$]  & $9.5\times10^{-2}$ & $6.4_{-0.8}^{+0.6}{~}_{-0.1}^{+0.0}\times10^{-2}$ \\
$N_{\rm hi}$\footnotemark[$\S$]  & -- & $3.5_{-0.5}^{+0.6}{~}_{-0.1}^{+0.1}\times10^{-2}$ \\
$\chi^2/{\rm d.o.f.}$ & 1139.6/1004 & 1035.4/1002 \\
%update 081024
\hline %-----------------------------------------------------------------
    \end{tabular}
  \end{center}
  \footnotesize
\noindent
\footnotemark[$*$] Temperature of the (cooler) thermal 
component in the $1kT$ ($2kT$) fit, in keV. 
Errors are 90\% statistical and systematic, respectively, 
latter evaluated by shifting the NXB and CXB within their systematic errors.\\
\footnotemark[$\dagger$] Temperature of the hotter thermal component in the $2kT$ fit.\\
\footnotemark[$\ddagger$] Metal abundance.\\
\footnotemark[$\S$] Normalization in the {\it apec} code, for the low and high temperature components.\\
\normalsize
\end{table}
%%%########################################################################

We first applied a single temperature thermal emission model (hereafter $1kT$ model)
to the XIS spectra, utilizing the  {\it apec} code in the XSPEC (v11.3.2) software package.
Galactic absorption was modeled by the {\it wabs} code with a fixed column 
density of $4.7\times10^{20}$ cm$^{-2}$, as derived by the radio observations
(Dickey \& Lockman 1990). %Dickey & Lockman, 1990, ARAA. 28, p215
The redshift was also fixed at $z = 0.0556$. 
Relative ratio of metal abundances are fixed to the solar value, given by Anders \& Grevesse (1989).
The temperature and the normalization were also set free.

Fit with the $1kT$ model was unsuccessful with $\chi^2/{\rm d.o.f.} = 1139.6/1004$,
leading to null hypothesis probability of $1.8\times10^{-3}$.
The fitted temperature was about 6.8 keV, which is generally consistent 
with temperatures given in the literature (e.g. Fukazawa et al. 2004).
Figure \ref{fig:center_xis_spec1} presents the data compared to the best fit model.
In the $1kT$ fitting, residuals around 1 keV and above 8 keV are significant.
Even if we set the $N_{\rm H}$ free, the fit did not improve significantly,
suggesting the thermal component modeling itself needs improvement.

Considering the multi-temperature nature of the cluster (e.g. Briel et al.~2004),
we added another thermal component.
Metal abundances of the two {\it apec} components were tied to be the same,
and only the temperature and normalization were set free (hereafter $2kT$ model).
The fit improved significantly and became acceptable 
with  $\chi^2/{\rm d.o.f.} = 1035.4/1002$ or null hypothesis probability of $23$\%.
Here, we also estimated the systematic error by
shifting both the NXB and CXB to their highest and lowest possible value at the 90\% confidence.
The shifts observed in the fitted central value are quoted as systematic uncertainty.
In this fitting, the hotter component temperature is suggested to be higher than 
17.1 keV. % sqrt
Here (and in all of the following analysis), the errors 
are derived as a quadrature sum of the statistical and
the systematic ones when obtaining the final result,
while the two components are separately presented in the tables for clarity.
When fitted using a model with a thermal and a power-law component (hereafter $1kT+PL$ model),
the fit also improves providing $\chi^2/{\rm d.o.f.} = 1057.8/1002$ or null hypothesis probability of $10.1$\%.
The photon index derived is hard with  $\Gamma = 1.0_{-0.8}^{+0.2}$. 
Before further modeling the ICM, we will check the HXD-PIN spectra
as it should contain informations about hotter or harder components.

\subsubsection{The HXD-PIN}
\label{chap:center_thermal_pin}

The signal detected by the PIN is significant compared to any possible
systematic errors. We then studied the shape of the PIN spectrum in the 13--40 keV range. 
Here, we used the HXD nominal position response, 
and did not use the arf files generated at section \ref{chap:arfgen} yet.
When we simply fit a power-law model to the PIN spectrum,
the resultant photon index $\Gamma$ is as soft as $3.6_{-0.9}^{+1.1}$.
If we try the $1kT$ model with metal abundances fixed at 0.3,
the resultant temperature is derived as $8.5_{-3.4}^{+5.8}$ keV.
Detailed results are presented in table \ref{table:pl_1kt_cenhxd}.
At the first glance, no strong signature of spectrally hard component is suggested in the PIN spectra.
We also note that, although the error bar is large because of both the statistical and systematic errors,
the photon index $\Gamma$ is softer than those suggested 
from the radio observation, i.e. $\Gamma \approx 2.1$ and the XIS spectral fitting with $1kT+PL$ model.
On the other hand, the PIN flux level provides important results,
as shown in the next section.

%%%######################### table 2-2-1 #######################################
\begin{table}
\caption{Spectral parameters fitted to the HXD spectra of the center pointing in the 13--40 keV range.}
\label{table:pl_1kt_cenhxd}
  \begin{center}
    \begin{tabular}{llll}
\hline %-----------------------------------------------------------------
\hline %-----------------------------------------------------------------
 & $\Gamma$ or $kT$\footnotemark[$*$]  & flux or norm\footnotemark[$\dagger$] & $\chi^2 / {\rm d.o.f.}$  \\
\hline %-----------------------------------------------------------------
PL model & $3.6_{-0.8}^{+1.0}{\ }_{-0.3}^{+0.5}$ & $32_{-5}^{+5}{\ }_{-6}^{+6}$  & 6.2/10 \\
$1kT$ model  & $8.5_{-2.9}^{+5.6}{\ }_{-1.8}^{+1.6}$ & $7.8_{-4.1}^{+12.1}{\ }_{-1.0}^{+2.2}$  & 6.2/10 \\
%% newest 081024
\hline %-----------------------------------------------------------------
\end{tabular}
\end{center}
\footnotesize
\footnotemark[$*$] Photon index $\Gamma$ or
thermal component temperature (in keV) of the power-law or $1kT$ fitting, respectively.\\
\footnotemark[$\dagger$] The 10-40 keV flux of the power-law component, in $10^{-12}$ erg s$^{-1}$ cm$^{-2}$,
or the normalization in {\it apec} model, for the low and high temperature components.
\normalsize
\end{table}
%%%########################################################################

\subsubsection{Combined XIS-HXD analysis}
\label{chap:center_thermal_wide}
\label{chap:center_2thermal}
\label{chap:center_nonthermal}

Following the spectral analysis individually 
performed to the XIS and the PIN,
in this section we give the results of a wide-band spectroscopic analysis,
using the two spectra combined (the XIS+PIN combined spectra).
In order to connect the two instruments and derive the ICM component,
we utilized the arf files generated in section \ref{chap:arfgen},
and fixed the normalization of the XIS-FI and the PIN spectra.

\begin{figure}
\begin{center}
    \FigureFile(6cm,8cm){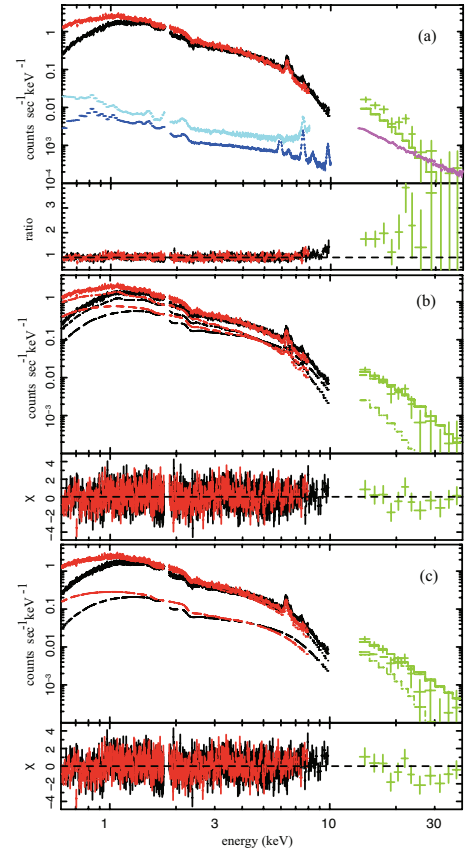} 
\end{center}
\caption{Wide-band spectra of the center pointing, fitted with the (a) $1kT$, (b) $2kT$
and (c) $1kT+PL$ model. In the panel (a), typical 90\% confidence systematic error in 
the NXB and CXB are also shown for reference, in cyan, blue and magenta
corresponding to the XIS-FI, BI and PIN, respectively.
Panel (a) is shown together with the residual ratio, 
while the other two are with the delta-chi distribution. }
\label{fig:center_spec1}
\end{figure}

We first applied the $1kT$ model. 
Procedures and fitting parameters are the same to those of the XIS-only analysis.
The fit again gave a temperature of $\sim 6.9$ keV.
However, as shown in figure \ref{fig:center_spec1} (a), the $1kT$ model failed to 
explain the PIN data, as well as the XIS data in the energy band above 7 keV.
As a result, the fit is unacceptable with $\chi^2/{\rm d.o.f.} = 1173.4/1016$. %081024
What is more, the observed PIN counts are about a factor of 2 larger than the best-fit $1kT$ model.
This excess is larger than any possible calibration errors in the arf,
and any systematic error associated with the NXB and CXB, as well.

%%#################### figure 1 #########################################
%%#######################################################################

To characterize the hard excess, we next applied the $2kT$ model.
Relative normalization of the two components
are the same in the XIS and the HXD-PIN spectra.
In other words, we assume a simple model 
in which the ICM emission uniformly contains the two components. 
Since 75\% of the ICM emission accumulated in the center PIN data
comes from within $12.5'$ from the X-ray centroid, which is mostly covered by the 
center XIS FOV, this model is valid as a simple approximation.
As shown in figure \ref{fig:center_spec1} (b) and table \ref{table:cenfit},
a good fit with $\chi^2/{\rm d.o.f.} = 1046.5/1014$ 
or a null-hypothesis probability of 23.3\% was obtained.
Interestingly, both the XIS and the PIN hard excess {\it consistently} disappeared.
Considering both the statistical and systematic errors (table \ref{table:cenfit}),
this hard component, if interpreted as thermal, is characterized by a temperature
of $kT_{\rm hi}$ ranging from 13.2--25.8 keV. %10.6--28.2 keV.

%%%######################### table #######################################
\begin{table}
\caption{The best fit parameters of the $2kT$ and $1kT+PL$ models
fitted to the XIS+PIN combined spectra of the center pointing.
Columns are similar to those shown in table \ref{table:cenfit_xis}.
The systematic error includes the XIS-PIN arf uncertainty of 10\%,
in addition to those of the CXB and NXB.
}
\label{table:cenfit}
\begin{center}
\begin{tabular}{lll}
\hline %-----------------------------------------------------------------
\hline %-----------------------------------------------------------------
 & $2kT$ & $1kT+PL$ \\
 \hline %-----------------------------------------------------------------
$kT_{\rm low}$ & $4.69_{-0.40}^{+0.29}{\ }_{-0.08}^{+0.01}$
	& $6.05_{-0.20}^{+0.18}{\ }_{-0.28}^{+0.24}$ \\
$kT_{\rm hi}$ & $19.2_{-4.0}^{+4.7}{\ }_{-4.5}^{+4.6}$ 
	& -- \\
$Z$& $0.39_{-0.03}^{+0.03}{\ }_{-0.01}^{+0.00}$
	& $0.35_{-0.03}^{+0.03}{\ }_{-0.01}^{+0.01}$\\
$N_{\rm lo}$ & $6.1_{-0.8}^{+0.6}{\ }_{-0.4}^{+0.2} \times 10^{-2}$ 
	& $8.2_{-0.4}^{+0.5}{\ }_{-0.3}^{+0.5} \times 10^{-2}$ \\
$N_{\rm hi}$ & $3.6_{-0.6}^{+0.8}{\ }_{-0.0}^{+0.4} \times 10^{-2}$ 
	& --\\
$\Gamma_{\rm PL}$\footnotemark[$*$]  & -- & $1.39_{-0.17}^{+0.10}{\ }_{-0.02}^{+0.00}$\\
$F_{\rm PL}$\footnotemark[$\dagger$]  & -- & $37_{-6}^{+7}{\ }_{-13}^{+14}$\\
$\chi^2/{\rm d.o.f.}$ & 1046.5/1014  & 1089.8/1014\\
% ALL NEWEST at 081024
\hline %-----------------------------------------------------------------
    \end{tabular}
  \end{center}
  \footnotesize
\noindent

\footnotemark[$*$] Photon index of the power-law component.\\
\footnotemark[$\dagger$] 10--40 keV flux of the power-law component,
in $10^{-12}$ erg s$^{-1}$ cm$^{-2}$.
\normalsize
\end{table}
%%%########################################################################

%%%######################### table #######################################
\begin{table}
\caption{The best fit parameters of the $3kT$ model
fitted to the XIS+PIN combined spectra of the center pointing.}
\label{table:cenfit_3kt}
\begin{center}
\begin{tabular}{ll}
\hline %-----------------------------------------------------------------
\hline %-----------------------------------------------------------------
 & $3kT$ \\
 \hline %-----------------------------------------------------------------
$kT_{\rm low}$ & $3.68_{-1.14}^{+0.56}{\ }_{-0.86}^{+0.00}$\\
$kT_{\rm med}$\footnotemark[$*$]  & $7.5$ (fixed)\\
$kT_{\rm hi}$ & $27.3_{-8.2}^{+\infty}{\ }_{-0.0}^{+\infty}$ \\
$Z$& $0.38_{-0.02}^{+0.02}{\ }_{-0.02}^{+0.00}$ \\
$N_{\rm lo}$ & $3.0_{-1.6}^{+1.5}{\ }_{-1.4}^{+0.0} \times 10^{-2}$ \\
$N_{\rm med}$ & $4.7_{-2.5}^{+2.5}{\ }_{-0.0}^{+2.3} \times 10^{-2}$ \\
$N_{\rm hi}$ & $2.1_{-1.1}^{+1.1}{\ }_{-1.1}^{+0.0} \times 10^{-2}$ \\
$\chi^2/{\rm d.o.f.}$ & 1046.5/1014  \\
% ALL NEWEST at 081024
\hline %-----------------------------------------------------------------
    \end{tabular}
  \end{center}
  \footnotesize
\noindent
\footnotemark[$*$] A fixed temperature of the medium hot component.
\normalsize
\end{table}
%%%########################################################################

As another candidate, we fitted the $1kT+PL$ model to the XIS+PIN combined spectra.
The fit was improved compared to the $1kT$ model,
and marginally acceptable giving 4.8\% in null hypothesis probability.
As shown in table \ref{table:cenfit}, the photon index was hard with $\Gamma = 1.4$
and the inferred power-law flux in 10--40 keV was $\sim 4\times10^{-11}$ erg s$^{-1}$ cm$^{-2}$,
normalized to the cluster as a whole.

The $2kT$ model is the simplest one to evaluate the
multi-temperature nature of the cluster ICM. 
Then we also tried a model with three {\it apec} components (hereafter $3kT$ model).
In this case, we fixed one of the three temperatures to 7.5 keV to
make the fit converge.
The value refers to the $1kT$ fit to the $17'$ offset data shown in the next section,
and is also consistent with one of the major components
seen in the {\it XMM} mosaic (Briel et al. 2004).
As shown in table \ref{table:cenfit_3kt},
the fit was acceptable with a null hypothesis probability of 27\%.
A hot component with a temperature $kT_{\rm hi} = 27.3_{-8.2}^{+\infty}$ keV
is required, but with $\sim 40$\% smaller normalization
compared to those of the hotter component of the $2kT$ model.
If we shift the temperature ranging from 6.5 to 8.0 keV,
still the fit is acceptable and the parameter value is generally similar.
In other words, we still need a component as hot as $\sim 20$ keV,
which could also be replaced by a $\Gamma \sim 1.4$ power law component.

The temperature map drived from {\it XMM} mosaic shows several
hot regions with $kT > 8$ keV around a few minutes 
north-west to the X-ray centroid (see figure 2 of Briel et al. 2004).
Using {\it Chandra}, the signature of a component as hot as $>10$ keV is also suggested
in the same region (Vikhlinin et al., private communication).
Due to projection of emission components, these value should be regarded 
as a lower limit, and thus a very hot ($> 13.2$ keV) component 
apparently is required in the cluster.
In the following analyses, we regard the $2kT$ model as an upper limit of the ICM contribution 
to the PIN-band, which is the most important issue in this paper.
Using additional long-exposure {\it XMM} data to be obtained soon,
we will further pursue this issue in the next paper.

Through these analyses,
we conclude that the excess hard component suggested in the XIS data
is clearly required in the PIN spectra as well.
The emission is likely to be of thermal origin, 
a very hot component with temperature exceeding 13.2 keV,
or less likely a hard ($\Gamma \sim 1.5$) power-law component.
We also checked the Fe-K line profile, and found that it was 
consistent with either the $1kT+PL$, $2kT$ and $3kT$ interpretations.
In any case, the analysis of the other observations, 
i.e. the 17$'$ offset and the NWR pointings, will be helpful to identify the origin of the hard emission.
The XIS and PIN data in these pointings contains spatial information of 
the excess component. What is more, the strong north-west radio relic 
is observed with the deep (81 ks in the XIS) NWR pointing.

\subsection{Spectral properties of the offset pointings}
\label{chap:offset_chk}

%%#################### figure 1 #########################################
\begin{figure}
\begin{center}
    \FigureFile(6cm,8cm){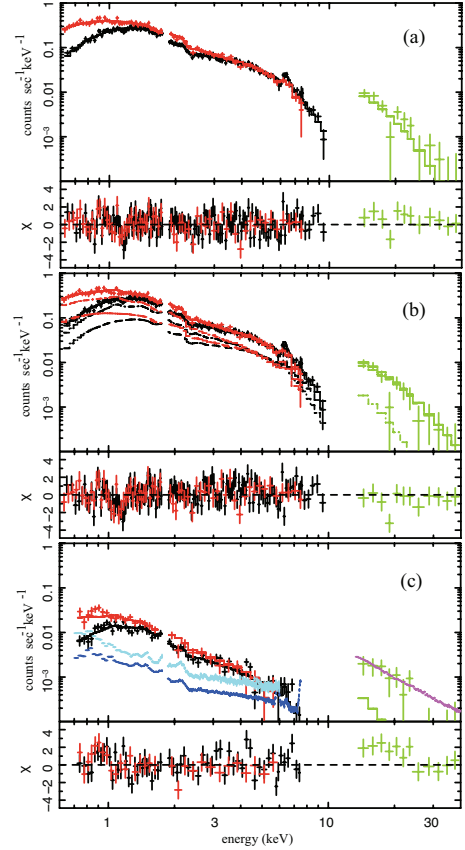} 
\end{center}
\caption{Wide-band spectra of (a) the $17'$ offset pointing fitted with $1kT$ model,
(b) plotted with $2kT$ best-fit model of the central pointing.
(c) The NWR spectra with the $1kT$ best-fit model to the XIS.
Cyan, blue and magenta plots show the
typical systematic error of the XIS-FIs, BI and the PIN spectra. See text for details.}
\label{fig:17_nwr_spec1}
\end{figure}
%%#######################################################################

%%%######################### table #######################################
\begin{table}
  \caption{Similar to table \ref{table:cenfit}, but for the $17'$ off and 
  the NWR offset spectra fitted with the $1kT$ model.}
  \label{table:offsetfits}
  \begin{center}
    \begin{tabular}{lll}
\hline %-----------------------------------------------------------------
\hline %-----------------------------------------------------------------
% & \multicolumn{2}{c}{Power law}  &  \multicolumn{2}{c}{2-kT thermal} \footnotemark[$\ddagger$] & chi$^2$/d.o.f.  \\
 & $17'$ off\footnotemark[$*$] & NWR\footnotemark[$\dagger$] \\
 \hline %-----------------------------------------------------------------
$kT$ [keV] & $7.71_{-0.32}^{+0.33}{\ }_{-0.47}^{+0.38}$ 
	& $3.6_{-0.4}^{+0.4}{\ }_{-0.6}^{+0.5}$ \\
$Z$ & $0.26_{-0.06}^{+0.06}{\ }_{-0.01}^{+0.01}$
	& $0.16_{-0.14}^{+0.16}{\ }_{-0.00}^{+0.01}$\\
$N$\footnotemark[$\ddagger$] & $9.3_{-0.2}^{+0.2}{\ }_{-0.2}^{+0.1} \times 10^{-2}$ &
	 $7.4_{-0.5}^{+0.6}{\ }_{-1.6}^{+1.2} \times 10^{-2}$ \\
$\chi^2/{\rm d.o.f.}$ & 234.1/220 & 120.5/89 \\
% NEWEST 081024
\hline %-----------------------------------------------------------------
    \end{tabular}
  \end{center}
  \footnotesize
\noindent
\footnotemark[$*$] Fitted to the XIS+PIN combined spectra.\\
\footnotemark[$\dagger$] Fitted only to the XIS spectra.\\
\footnotemark[$\ddagger$] Normalization in {\it apec} model. Note that it is normalized to the whole cluster, not within the observation FOV, using the arf files generated from the PSPC image.
\normalsize
\end{table}
%%%########################################################################

When analyzing the offset pointings, we have to take into account that,
although the XIS spectra are of local origin (from within the FOV),
the PIN spectra is largely contaminated from the emission from the cluster center.
For example, 59\% and 37\% of the ICM emission
accumulated in the  $17'$offset and the NWR pointings, respectively,
comes from within $12.5'$ from the X-ray centroid, i.e. around the center.
Therefore the XIS and the PIN spectra wil not necessarily show the same results,
although they should be roughly consistent with one another as a whole.

The XIS spectra of the 17$'$ offset pointing can be represented using 
the $1kT$ model, with $kT = 7.6_{-0.4}^{+0.4}$ keV and $\chi2/{\rm d.o.f.} = 221.9/208$, %081024
or a null hypothesis probability of 24.3\%.
Note that we utilized the arf files generated in section \ref{chap:arfgen} 
so that the (energy dependent) vignetting effect is fully taken into account.
For the same reason, the fitted flux is normalized 
to the reference image used in the arf generation (in this case the whole cluster), 
not just the flux within the FOV.
When we use the XIS+PIN combined spectra,
the $1kT$ model is acceptable as well, with a null-hypothesis probability of 24.5\%.
Results of the fitting are presented in table  \ref{table:offsetfits}.
As shown in figure \ref{fig:17_nwr_spec1} (a), however, 
the model slightly (by $20$\%) under-predicts the data in the PIN band.
The residual resides right on the 90\% upper-limit of the CXB and NXB systematic error.

The 17$'$ offset data is also consistent with the $2kT$ picture of the center pointing.
In fact, when we applied the $2kT$ best fit model of the center pointing to the XIS-PIN spectra,
a marginally acceptable fit was obtained by shifting its global normalization 
only by $-3$\% (see also panel (b) of figure \ref{fig:17_nwr_spec1}).
The fit gives $\chi2/{\rm d.o.f.} = 260.8/222$ or a null-hypothesis probability of 3.8\%.
In this case, the model slightly (by $20$\%) over-predicts the PIN data.
One possibility is that the cluster as a whole can be represented by
the $2kT$ model, if one includes the systematic errors.
The result can also be understood if the very hot component resides 
within the center XIS FOV, and the majority of the $17'$ offset XIS FOV is
covered by a single $\sim 7.5$ keV thermal component.

In both the XIS and PIN spectra of the NWR pointing, the CXB and NXB systematic errors
are not negligible, as shown in the panel (c) of figure \ref{fig:17_nwr_spec1}.
Here,  we used the southern 1/3 region of the NWR pointing in the XIS analysis
(hereafter 1/3 NW-south region) because
the systematic errors make spectral analysis difficult in the other regions.
We restrict our fitting to the energy band from 0.7--7.5 keV and
0.7--6.0 keV in the XIS-FI and BI spectra, respectively.
We first applied the $1kT$ model to the XIS spectra.
The best-fit temperature thus obtained is as low as $3.6_{-0.7}^{+0.6}$ keV %lin {-1.0}^{+0.9}
(see table \ref{table:offsetfits}),
which is significantly cooler than those of the center and 17$'$ offset data.
The fit is marginally acceptable with a null-hypothesis probability of 1.5\%.
Noticeably large residuals appears around 0.9 keV suggesting 
an additional $\sim 1$ keV thermal emission component.
The XIS data around the north-west radio relic is further analyzed in section \ref{chap:nwr_xis}.

The NWR PIN signal could be explained by the CXB and NXB systematic errors, 
since the level is right on their 90\% confidence upper-limit (see the panel (c) of figure \ref{fig:17_nwr_spec1}).
Another reasonable possibility is the ICM contribution from the cluster central region to the PIN spectra.
Based on the arf files derived in section \ref{chap:arfgen},
37\% and 49\% of the ICM contribution to the NWR-PIN data is originated from
within 12.5$'$ and 26$'$ from the cluster X-ray centroid, respectively,
which roughly correspond to the area of the center and 17$'$ offset XIS FOVs.
%In other words, only 14\% is estimated to come
%from the cluster periphery, i.e. $>26'$ from the cluster center.
As shown later in section \ref{chap:nwr_pin_xis}, 
the ICM contribution estimated using the XIS 
can explain all of the PIN signals in the NWR pointing.
As a cross check, we also introduce the  {\it XMM} mosaic data (e.g. Briel et al. 2004)
covering the entire PIN FOV later in section \ref{chap:nwr_pin_xmm},
and come to the same conclusion.

In table \ref{table:pl_offsetshxd}, we present the PIN spectral properties separately fitted
to the 17$'$ offset and the NWR data. Using the power-law model,
the photon index $\Gamma$ becomes as soft as $3.5$.
Although the error is large, the best-fit value is consistent with those
obtained from the center PIN spectra (see table \ref{table:pl_1kt_cenhxd}),
while the flux itself is much brighter there.
These results strongly suggest that the same emission dominant in the center-PIN spectra
is contaminating both the 17$'$ offset and the NWR PIN data.

%%%######################### table 2-2-1 #######################################
\begin{table}
\caption{Spectral parameters of the power-law model 
fitted to the HXD spectra of the 17$'$ offset and the NWR pointings in the 13--40 keV range.}
\label{table:pl_offsetshxd}
  \begin{center}
    \begin{tabular}{llll}
\hline %-----------------------------------------------------------------
\hline %-----------------------------------------------------------------
Pointings & $\Gamma$ & flux\footnotemark[$*$]  & $\chi^2 / {\rm d.o.f.}$  \\
\hline %-----------------------------------------------------------------
17$'$ offset & $3.4_{-1.0}^{+1.2}{\ }_{-0.5}^{+1.0}$ & $20_{-4}^{+4}{\ }_{-6}^{+6}$ & 8.4/10 \\
NWR & $3.4_{-1.5}^{+2.0}{\ }_{-0.9}^{\rm N/A}$\footnotemark[$\dagger$] & $5.1_{-1.7}^{+1.7}{\ }_{-5.1}^{+6.3}$ & 7.5/10 \\
%081024
\hline %-----------------------------------------------------------------
\end{tabular}
\end{center}
\footnotesize
\footnotemark[$*$] 10-40 keV flux of the power-law component, in $10^{-12}$ erg s$^{-1}$ cm$^{-2}$.\\
\footnotemark[$\dagger$] 
N/A means the value was not available. In this case, the lack of photon made it impossible to restrict the spectral shape.
\normalsize
\end{table}
%%%########################################################################

\subsection{Joint fit to the wide-band spectra of the three pointings}
\label{chap:3-wide}
\label{chap:3-wide-2kT}

The PIN data of the three pointings are all mixed to some extent,
mostly contaminated from the emission around the cluster center.
In this section, we combine them to obtain the cluster overall properties.

Analysis of the three pointings suggests that all the data are consistent with
the simple two component picture suggested in the center pointing data,
except for the XIS data in the NWR pointing showing rather low ICM temperature. 
Because of its low surface brightness,
thermal contribution from this region is minor in the PIN band.
Thus, we conducted a spectral analysis to the data combining in total 5 data sets
(hereafter all-combined data),
i.e. three PIN spectra and the XIS spectra of the center and 17$'$ offset pointings.
Following the strategy employed at section \ref{chap:center}, 
we fixed the relative normalization between the center XIS-FI spectra
and all the PIN spectra, while those between the XISs (offset FI data and all the BI data)
were set free to handle small ($<5$\%) residual uncertainties. 
As noted, the 10\% systematic error between the XIS-FI and the PIN data
were handled separately.

As was the case with the center pointing, 
the $1kT$ model fitting to the all-combined spectra was unsuccessful,
giving $\chi^2/{\rm d.o.f.} = 1430.7/1250$ or a null-hypothesis probability of $3\times10^{-4}$. %081024
Thus, we applied the $2kT$ model to the all-combined spectra.

\begin{figure}
    \FigureFile(7.5cm,5cm){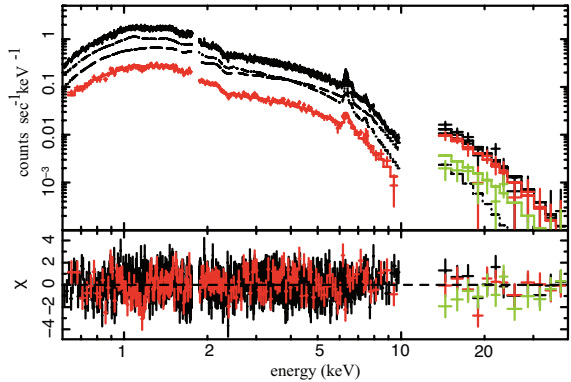} 
\caption{
The all-combined spectra fitted with the $2kT$ model. 
For simplicity, the XIS-BI spectra are not shown, although used in the actual fitting. 
The XIS and the PIN spectra of the center pointing are shown with black crosses, 
and those of the 17$'$ offset pointing in red. Green crosses give the PIN spectrum of the NWR pointing. 
Two components of the model is presented individually as a dashed line only for the
center data. }
\label{fig:3all_2kt_all}
\end{figure}

The $2kT$ fit was almost acceptable as shown 
in figure \ref{fig:3all_2kt_all} and table \ref{table:5allfit}.
The null-hypothesis probability of the fit was 7.6\%. %081024
Partly because of the slight discrepancy of the spectral shape 
around 1 keV between the center and the 17$'$ offset poitings,
the hotter component temperature was not determined well
when the NXB, the CXB and the arf systematic errors were included in the fit.
However, the best-fit value of the hotter component temperature $kT_{\rm hi}$ is still
14.7 keV, which is consistent with the results from the center pointing data
(see section \ref{chap:center_2thermal}). The $3kT$ model gives acceptable fit, as well.

The $1kT+PL$ model fitting to the all-combined spectra
gave results similar to those of the center pointing (table \ref{table:5allfit}).
The fit itself was marginally acceptable with
 a null-hypothesis probability of 1.2\%, significantly worse than that of the $2kT$ model fitting.
Although the poor fit could be due to an overly simplified model,
the fact that the HXD spectra have a shape as soft as $\Gamma = 3.5$ strongly 
indicates emission of thermal origin.
Thus here we conclude that the all-combined fitting also prefers the $2kT$ (or $3kT$) picture.
Another picture that the very hot component resides only around the cluster center
may apply as well, considering the systematic errors. In any case,
the NWR-PIN data is well explained by 
the contribution from the cluster central portions, 
suggesting there is no strong signal of the IC emission in the data.

%%#################### figure 2-2-1 #########################################
%%#######################################################################

%%%######################### table #######################################
\begin{table}
\caption{Best fit parameters of the $2kT$ model and $1kT+PL$ model fitted to the all-combined spectra. 
Columns are similar to those presented in table \ref{table:cenfit}. }
\label{table:5allfit}
\begin{center}
\begin{tabular}{lll}
\hline %-----------------------------------------------------------------
\hline %-----------------------------------------------------------------
% & \multicolumn{2}{c}{Power law}  &  \multicolumn{2}{c}{2-kT thermal} \footnotemark[$\ddagger$] & chi$^2$/d.o.f.  \\
%Model & $2kT$ \\ update 070802 from 070716r_3all_2vapec_b1_2.xcm
& $2kT$ &	 $1kT+PL$ \\
\hline
$kT_{\rm lo}$ &	 $4.72_{-0.60}^{+0.36}{\ }_{-2.81}^{+0.01}$ &
	$6.40_{-0.16}^{+0.15}{\ }_{-0.43}^{+0.32}$\\
$kT_{\rm hi}$  & $14.7_{-3.2}^{+3.1}{\ }_{-7.2}^{+10.4}$ &
	-- \\
$Z$ & 	$0.38_{-0.02}^{+0.02}{\ }_{-0.04}^{+0.00}$ &
	$0.34_{-0.02}^{+0.02}{\ }_{-0.02}^{+0.02}$\\
$N_{\rm lo}$  & $5.6_{-1.4}^{+0.8}{\ }_{-5.0}^{+0.8}\times 10^{-2}$ &
	$8.5_{-0.3}^{+0.4}{\ }_{-0.8}^{+0.8}\times 10^{-2}$ \\
$N_{\rm hi}$  & $4.1_{-0.8}^{+1.3}{\ }_{-1.6}^{+4.8}\times 10^{-2}$ &
	-- \\
$\Gamma_{\rm PL}$ & -- & $1.46_{-0.16}^{+0.10}{\ }_{-0.02}^{+0.01}$\\
$F_{\rm PL}$ & -- & $25.1_{-4.8}^{+5.3}{\ }_{-21.7}^{+22.5}$\\
$\chi^2/{\rm d.o.f.}$ & $1320.4/1248$ &	$1364.0/1248$\\
\hline %-----------------------------------------------------------------
%081024
    \end{tabular}
  \end{center}
  \footnotesize
\noindent
%\footnotemark[$*$] Constant factor for the $17'$ offset XIS-FI and 
%the center, $17'$ offset and NWR PIN spectra normalized to the center XIS-FI data.\\
\normalsize
\end{table}
%%%########################################################################

\subsection{Search for the IC signal from the north-west radio relic}
\label{chap:NWRall}

The north-west radio relic region is mostly covered by
the deep (81 ks in the XIS and 48 ks in the HXD) NWR pointing.
Although there is no strong evidence for the IC emission in the analysis so far,
there should be some X-ray and hard X-ray emission from the relic.
In this section, we focused mainly on this region 
to detect or determine the upper-limit on the IC emission.

\subsubsection{The HXD upper limit combined with the XIS results}
\label{chap:nwr_pin_xis}

%%%######################### table #######################################
\begin{table}
\caption{Event counts in the 13--40 keV NWR-PIN data, compared to various error components.}
\label{table:nwr_count_pin}
\begin{center}
\begin{tabular}{ll}
\hline %-----------------------------------------------------------------
\hline %-----------------------------------------------------------------
Components	& counts s$^{-1}$\\
\hline
Signal & $({14.1}\pm{4.9})\times10^{-3}$ \\  % 2.985*1.645 = 4.91
4.5\% of NXB\footnotemark[$*$]  & $\pm {15.0}\times10^{-3}$ \\
18\% of CXB\footnotemark[$\dagger$]  & $\pm {4.8}\times10^{-3}$ \\
ICM$_{2kT}$\footnotemark[$\ddagger$] & $({28.4}{~}\pm{3.0}\pm{2.8})\times10^{-3}$ \\
ICM$_{1kT}$\footnotemark[$\ddagger$] & $({16.0}{~}\pm{2.0}\pm{1.6})\times10^{-3}$ \\
ICM$_{{\it XMM}}$\footnotemark[$\ddagger$] & $({14.7}{~}\pm{1.5}{~}\pm{1.5})\times10^{-3}$ \\
\hline %-----------------------------------------------------------------
Signal$_{2kT}$\footnotemark[$\S$] & $({-14.3}\pm{4.9}\pm{16.3})\times10^{-3}$ \\ 
Signal$_{1kT}$\footnotemark[$\S$] & $({-1.9}\pm{4.9}\pm{16.0})\times10^{-3}$ \\ 
Signal$_{{\it XMM}}$\footnotemark[$\S$] & $({-0.6}\pm{4.9}\pm{15.9})\times10^{-3}$ \\ 
\hline %-----------------------------------------------------------------
$F_{2kT~{\rm PL}}$\footnotemark[$\#$] & ${-5.1}\pm{1.8}\pm{5.8}$ \\  
$F_{1kT~{\rm PL}}$\footnotemark[$\#$] & ${-0.7}\pm{1.8}\pm{5.7}$ \\  
$F_{{\it XMM}~{\rm PL}}$\footnotemark[$\#$] & ${-0.2}\pm{1.8}\pm{5.7}$ \\  %2.80e-3 cts/s is 1e-12 10-40 keV
\hline %-----------------------------------------------------------------
%081024
    \end{tabular}
  \end{center}
  \footnotesize
\noindent
\footnotemark[$*$] Estimated NXB systematic error at the 90\% confidence.\\
\footnotemark[$\dagger$] Estimated CXB fluctuations at the 90\% confidence.\\
\footnotemark[$\ddagger$] ICM emission component estimated using the 
$2kT$ model fitted to the center XIS-PIN combined data,
the $1kT$ model fitted to the 17$'$ offset XIS data, and to the {\it XMM} weighted data.
Values are shown with 90\% fitting error including the NXB+CXB systematic effect in quadrature sum
and 10\% arf calibration systematic error. \\
\footnotemark[$\S$] Non-thermal signals in counts s$^{-1}$ with 90\% statistical and systematic errors. 
The latter value is the quadrature sum of those associated with NXB, CXB and the ICM.\\
\footnotemark[$\#$] The 10--40 keV band flux in
$10^{-12}$ erg s$^{-1}$ cm$^{-2}$ derived from the ``Signal'' column,
converted to a $\Gamma = 2.1$ power law emission.
\normalsize
\end{table}
%%%########################################################################

We evaluated the flux of the possible non-thermal hard X-ray emission
using the 13--40 keV PIN signal of the NWR pointing.
In table \ref{table:nwr_count_pin}, signal rates are summarized with their errors.
To estimate the ICM components, we used 
both the $2kT$ model fitted to the center XIS-PIN combined spectra
and the $1kT$ model fitted to the 17$'$ offset XIS-only spectra.
The 17$'$ offset PIN data was not used, 
because it may contain some signal from the north-west radio relic.

Using the center-$2kT$ model, we converted the residual signal in the NWR-PIN data 
by assuming a $\Gamma = 2.1$ fixed power-law emission located at the FOV center.
The result is $({-5.1}\pm{1.8}\pm{5.8}) \times 10^{-12}$ erg s$^{-1}$ cm$^{-2}$,
i.e. $< 1.0 \times 10^{-12}$ erg s$^{-1}$ cm$^{-2}$ in the 10--40 keV band.
Here again, the CXB and the NXB systematic errors
and the 10\% systematic error in the arf files are taken into account.
Note that the arf files are used to estimate the ICM contribution,
but not in the hard excess flux evaluation.
Using the $1kT$ model fitted to the 17$'$ offset XIS spectra, the power-law flux is estimated to be
$({-0.7}\pm{1.8}\pm{5.7}) \times 10^{-12}$ erg s$^{-1}$ cm$^{-2}$, 
i.e. $< 5.3 \times 10^{-12}$ erg s$^{-1}$ cm$^{-2}$ in the 10--40 keV band. %lin 7.0
Since the contribution from the central portion and 17$'$ offset region are
both significant, actual upper-limit on the power-law emission will be
somewhere between these two values.
Here, we recognize the latter results, $< 5.3 \times 10^{-12}$ erg s$^{-1}$ cm$^{-2}$, %lin 7.0
as the conservative upper-limit flux of the excess hard X-rays from the north-west relic.

\subsubsection{The HXD upper limit combined with the {\it XMM} results}
\label{chap:nwr_pin_xmm}

%%#################### figure 1 #########################################
\begin{figure}
\begin{center}
    \FigureFile(7.5cm,7.5cm){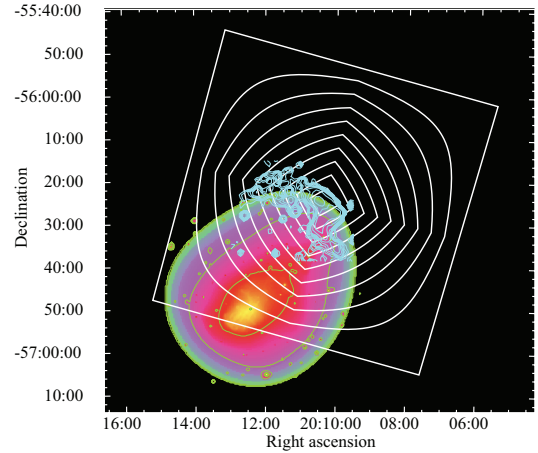} 
\end{center}
\caption{The wavelet smoothed {\it XMM} mosaic image
of Abell 3667 with a logarthmic color scale (Briel et al. 2004).
The white square is the total FOV of the PIN for the NWR
observation.  The interior white regions are curves of constant
PIN effective area, ranging from 10\% to 90\% of the effective
area for a central point source.  
The cyan contours are from the SUMSS 842 MHz radio image of the
NW radio relic. See text for detail.}
\label{fig:xmm-pin_nwr_im}
\end{figure}
%%#######################################################################

%%#################### figure 1 #########################################
\begin{figure}
\begin{center}
    \FigureFile(7.5cm,5cm){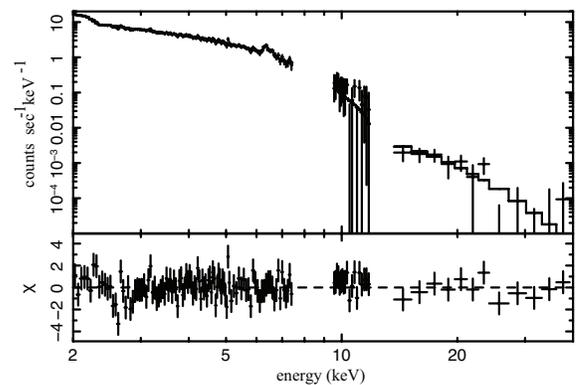} 
\end{center}
\caption{The NWR-PIN spectra combined with the {\it XMM} spectra. The best-fit $1kT$ model is shown.}
\label{fig:xmm-pin_nwr}
\end{figure}
%%#######################################################################

%%%######################### table #######################################
\begin{table}
\caption{Best fit parameters of the $1kT$ model fitted to the {\it XMM}-pn spectra,
weighted with the HXD-PIN vignetting function of the NWR pointing.}
\label{table:xmmfit}
\begin{center}
\begin{tabular}{lll}
\hline %-----------------------------------------------------------------
\hline %-----------------------------------------------------------------
$kT$\footnotemark[$*$]  &	Abun.\footnotemark[$\dagger$]   & $\chi^2/{\rm d.o.f.}$\\
\hline
$7.93_{-0.42}^{+0.43}$ & 	$0.22_{-0.04}^{+0.04}$ & 116.6/123 \\
\hline %-----------------------------------------------------------------
    \end{tabular}
  \end{center}
  \footnotesize
\noindent
\footnotemark[$*$] Temperature of the {\it apec} model.\\
\footnotemark[$*$] Metal abundance of the {\it apec} model.
\normalsize
\end{table}
%%%########################################################################

Mapping observations performed by {\it XMM}, as described in \citet{BFH04},
provides independent information of the ICM and its temperature distribution.
With its larger FOV, the {\it XMM} data has a merit on observing 
almost all of the ICM emission covered by the HXD-PIN FOV,
while the XIS has the merit of low background and simpler cross-calibration with the HXD.

The {\it XMM} spectra, background and arf files for this analysis are generated 
using the method described in Wik et al. (2008). 
In summary, we defined 10 regions 
with nearly constant PIN effective area, separated by curves with
0\%, 10\%, 20\%, ... 90\% of the value for the central point source,
as shown in figure \ref{fig:xmm-pin_nwr_im}.
Then we extracted {\it XMM} EPIC-pn spectra from these region in the standard manner,
using the {\it XMM}-mosaic data and adopting the XMMSAS v7.1 calibration.
The background spectrum is derived from the datasets compiled by \citet{RP03},
and we also apply consistent flare cleaning criteria to both the data and the background.
The arf files used for the XMM data include the corrections for exposure and vignetting for the XMM EPIC-pn.

By adding the 10 signal spectra multiplied by the PIN
vignetting function, and correcting the XMM-XIS cross
normalization of 15\%,
we obtain the XMM spectrum that would be observed with
XMM but if it had the spatial spatial response as the Suzaku PIN detector.  
The normalization of the spectrum is based on the response of the PIN to 
a point source at the center of the PIN field of view; thus, the fluxes are correct
if the PIN spectra are fit using the arf appropriate for a central point source.
To avoid the contribution of the soft emission components to the 
results, we performed the spectral fitting above 2 keV, and also masked 
the energy-region in which the background lines (mostly from Ni) are prominent.

In figure \ref{fig:xmm-pin_nwr}, we present the {\it XMM}-pn spectra 
fitted with the $1kT$ model. Parameters are summarized in table~\ref{table:xmmfit}.
The NWR-PIN spectra is remarkably consistent with 
the {\it XMM}-best-fit model superposed, as also shown in the figure.
Thus, using the {\it XMM}, we independently obtained a result that 
no additional hard X-ray component is required in the NWR-PIN data.
As shown in table~\ref{table:nwr_count_pin},
the power-law flux is estimated to be
$({-0.2}\pm{1.8}\pm{5.7}) \times 10^{-12}$ erg s$^{-1}$ cm$^{-2}$, 
i.e. $< 5.8 \times 10^{-12}$ erg s$^{-1}$ cm$^{-2}$ in the 10--40 keV band. %lin 7.1

\subsubsection{The XIS image and spectral analysis}
\label{chap:nwr_xis}

%%#################### figure 2-2-1 #########################################
\begin{figure}
\begin{center}
    \FigureFile(7.5cm,7.5cm){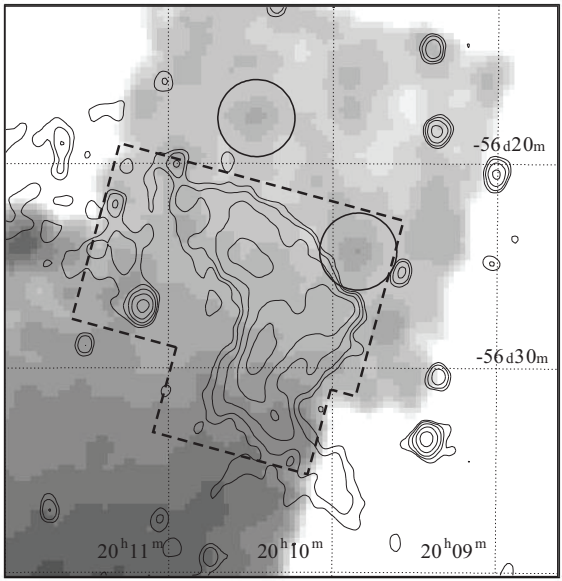} 
\end{center}
\caption{The XIS 1--4 keV mosaic image (gray scale) with 843 MHz image overlaid (contours).
Regions $2'$ around the two point sources are shown as solid circles.
The north-west relic region, used in the spectral analysis, is shown with thick dashed line.}
\label{fig:a3667_xis_nwr_image}
\end{figure}
%%#######################################################################

%%#################### figure 2-2-1 #########################################
\begin{figure}
\begin{center}
    \FigureFile(7.5cm,5cm){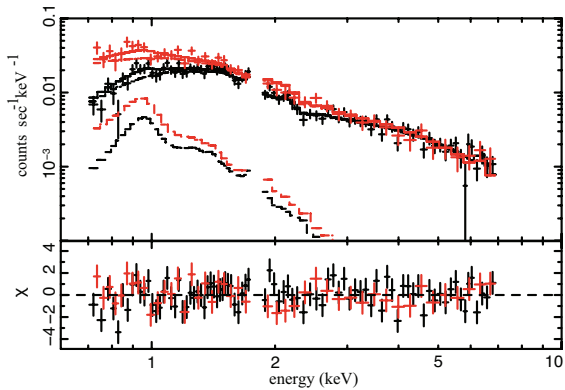} 
\end{center}
\caption{The XIS spectra of the north-west relic region fitted with $2kT$ thermal model. 
Note the cool component at $\sim 0.9$~keV.}
\label{fig:a3667_xis_nwr_spec}
\end{figure}
%%#######################################################################

Most of the north-west radio relic is covered by the XIS FOVs
of the 17$'$ offset and the NWR pointings.
Since the background of the XIS is low and stable,
the data contains good informations on the IC emission from the relic itself.
We then analyzed the XIS spectra obtained from the 
region as shown in figure \ref{fig:a3667_xis_nwr_image}, 
which covers $90$\% of the 847 MHz radio flux (hereafter the north-west relic region).
We also examined the projected brightness profile in three energy bands,
i.e. 1--2 keV, 2--4 keV and 4--8 keV, to search for any signal of IC emission.

In the north-west relic region and areas further north, 
the diffuse X-ray emission is weak and 
dimmer point sources can be identified easily.
As shown in figure \ref{fig:a3667_xis_nwr_image}, we identified two sources
with flux $1.5\sim 3.5\times 10^{-14}$ erg s$^{-1}$ cm$^{-2}$ in the 2--10 keV band,
and masked out $2'$ around them.
When we perform the same operation to the Lockman Hole observation, 
we found two sources as well. There the overall CXB level decreases by $\sim 10$\% 
after masking these sources.
This shift is fed back into the CXB estimation of the NWR analysis here.
Since the upper cutoff flux is changed to $S_c \sim 2\times10^{-14}$ erg s$^{-1}$ cm$^{-2}$
from $\sim 1\times 10^{-13}$ erg s$^{-1}$ cm$^{-2}$,
the CXB fluctuation level is reduced to 15.9\% from the original 21\%
(see also appendix \ref{apdx:xis_cxb}).

In figure \ref{fig:a3667_xis_nwr_spec}, the XIS-FI and BI spectra of the 
north-west relic region data obtained after masking the sources are shown.
Both the NWR and 17$'$ offset data were added.
To estimate the flux associated with the relic, 
we calculated the detector response to a widely distributed flat emission
accumulated within the north-west relic region, and then scaled it to the sky coverage
($\Omega_e = 0.04$ deg$^2$) of the region.
When fitted with a $1kT$ model, the fit is marginally acceptable, 
giving $\chi^2/{\rm d.o.f.} = 158.1/127$ or 3.2\% in null hypothesis probability.
The temperature is $5.8 \pm 0.7$ keV, 
between the two values obtained in the XIS data of the two pointings 
(see section \ref{chap:offset_chk}). 
To handle the significant hump at around 0.9 keV, 
we added another thermal component (the $2kT$ model).
Then we obtained almost acceptable fit with $\chi^2/{\rm d.o.f.} = 145.9/125$ or
9.7\% in null hypothesis probability.
The hot component temperature is derived as $kT_{\rm hi} = 7.0_{-1.6}^{+2.0}$ keV 
and that of the cool one as $kT_{\rm lo} = 1.03_{-0.92}^{+0.24}$ keV, 
while the common abundance is $0.17_{-0.12}^{+0.14}$ solar. 
The significance of the Fe-K lines is weak,
which is not surprising considering the low surface brightness.
Since the flux of the cool component at $\sim 0.9$ keV is similar to that of the CXB model,
the emission could be either a slight hump in the ``foreground Galactic component'' or
some cool component really associated with the cluster periphery.

%%%######################### table 2-2-1 #######################################
\begin{table}
\caption{Spectral parameters of the power-law model fitted to the XIS spectra of the north-west relic region .}
\label{table:pl_1/3nwr}
  \begin{center}
    \begin{tabular}{lll}
\hline %-----------------------------------------------------------------
\hline %-----------------------------------------------------------------
$\Gamma$ & flux\footnotemark[$*$]  & $\chi^2 / {\rm d.o.f.}$  \\
\hline %-----------------------------------------------------------------
$1.8_{-0.1}^{+0.1}{\ }_{-0.1}^{+0.1}$ & $0.62_{-0.10}^{+0.12}{\ }_{-0.19}^{+0.20}$ & 65.1/66 \\
\hline %-----------------------------------------------------------------
\end{tabular}
\end{center}
\footnotesize
\footnotemark[$*$] 10-40 keV flux of the power-law component, in $10^{-12}$ erg s$^{-1}$ cm$^{-2}$.\\
\normalsize
\end{table}
%%%######################### table 2-2-1 #######################################

Since the region coincides in position with the diffuse radio emission,
the detected X-rays might be of non-thermal origin.
As an extreme case, we fit the spectra with a power-law model.
Because of the 0.9 keV hump, which indicates a contribution from $kT \sim 1$ keV 
thermal component, we utilized the energy band above 1.5 keV in this fitting. 
We also limit it to below 6.0 keV since the NXB systematic error dominates above this energy.
The results are listed in table \ref{table:pl_1/3nwr}.
The power-law fit is acceptable with a null-hypothesis probability of 51\%.
Taking into account the large systematic errors associated with the NXB and CXB 
(see figure \ref{fig:17_nwr_spec1}c),
we cannot distinguish whether the emission is thermal or non-thermal from spectroscopy only.

\begin{figure}
\begin{center}
    \FigureFile(7.5cm,7.5cm){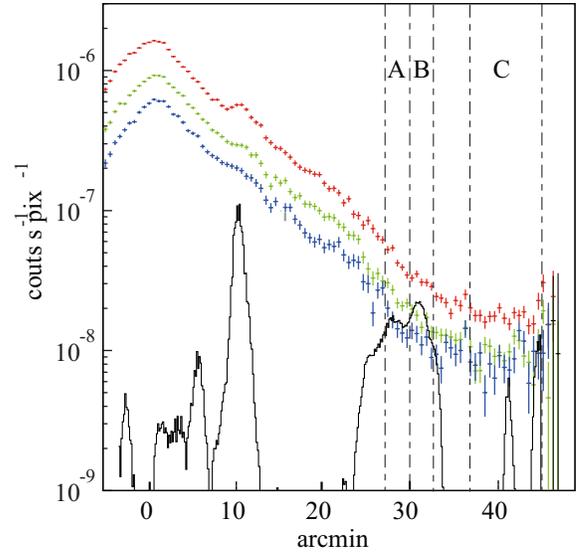} 
\end{center}
\caption{Projected X-ray image in the energy band of 1--2 keV, 2--4 keV
and 4--8 keV in red, green and blue crosses, respectively.
The width of the projection region is $12'$ along the merger axis, as shown in figure \ref{fig:xis_mosaic}.
Black histogram stands for SUMSS 843 MHz radio image obtained via Skyview service by NASA,
projected in the same manner but with arbitral normalization.
The region A, B and C are those used for IC upper-limit analysis. See text for detail.
}
\label{fig:a3667_projection}
\end{figure}

In figure \ref{fig:a3667_projection}, we present the projected X-ray count-rate map
in the energy bands of 1--2 keV, 2--4 keV and 4--8 keV derived from all the three 
XIS-FI detectors. Here, the BI chip is not utilized because of its higher background.
Regions around the two point source in the NWR data are excluded in the plot.
Overlaid is the profile of the SUMSS 843 MHz radio image, which clearly shows the relic,
located $\sim 30'~(2$ Mpc) away from the X-ray centroid.
From the 1--2 keV and 2--4 keV profiles, it is clear that the cluster emission was detected 
out to the region $40'$ from the cluster X-ray centroid, which corresponds to 2.6 Mpc in projected distance.
On the other hand, there is apparently no excess signal associated with the radio relic,
smoothly connecting the inner ICM X-rays to the CXB component at the outermost region.
Thus, a significant fraction of the X-rays in the NWR region was naturally assumed to be of thermal origin.

Using this plot, we roughly evaluated the upper-limit of the 
projected non-thermal count rates associated with the relic.
Here, we assumed that the IC emission has the same morphology to that of the radio emission,
and the 2--8 keV X-ray counts within region B in figure \ref{fig:a3667_projection} are the 
upper-limit to the IC emission. 
The CXB is estimated from region C, and its fluctuation is calculated to be 22\% on both the region A and B.
Taking the statistical error and the CXB fluctuation into account,
relative count-rate of region B to that of region A is
derived as $0.48_{-0.17}^{+0.22}$ while that of the radio flux is $1.17$.
Thus, we concluded that $40_{-16}^{+16}$\%, i.e. larger than 24\%,
of the emission in region A+B is of thermal origin.
Since the region A+B coincides with the norht-west relic region,
we combine this result with the power-law fitting results (table \ref{table:pl_1/3nwr}).
In addition, because the radio emission within the spectral fitting region 
only covers 90\% of the total relic flux in the SUMSS image, 
we multiply the XIS flux by a factor of 1.1 to account for any missing nonthermal flux.
Finally, we obtain the 10--40 keV upper limit IC flux for the north-west radio relic as a whole
as  $7.3\times10^{-13}$ erg s$^{-1}$ cm$^{-2}$ allowing
$\Gamma$ to vary, at the 90\% confidence level.
With this flux, the photon index is required to be $\Gamma = 1.6$.
If we fix it to $\Gamma = 2.1$, 
the 10--40 keV upper limit flux become $3.9\times10^{-13}$ erg s$^{-1}$ cm$^{-2}$.

%%#################### figure 2-2-1 #########################################
%%#######################################################################

%%%5. Discussion %%%%%%%%%%%%%%%%%%%%%%%%%%%%%%%%%%%%%%
\section{Discussion}
\label{chap:discussion}
%%%%%%%%%%%%%%%%%%%%%%%%%%%%%%%%%%%%%%%%%%%%%%%%%%%%%%%

\subsection{Short summary of the analysis results}
\label{chap:discussion-sum}

We analyzed the {\it Suzaku} mapping observation data of a merging cluster Abell~3667.
The XIS detects emission from the cluster out to $\sim 40'$ or 2.6 Mpc from the cluster X-ray centroid.
The PIN spectra of the three pointings commonly 
prefer soft-shaped spectra, suggesting that 
the thermal contribution from the cluster ICM is the major origin of these signals.
From the XIS+PIN combined analysis of the center pointing,
a multi-phase picture such as the $2kT$ model, the $1kT+PL$ model or the $3kT$ model is suggested.
In the former case, $kT_{\rm hi}$ is derived to exceed 13.2 keV, %10.6
with the best fit value around 20 keV, while $kT_{\rm lo}$ is around $4.7$ keV.
Using the $3kT$ model with one of the temperature fixed at 7.5 keV, 
the fit is also acceptable, while a component wiht $kT_{\rm hi} \sim 20$ keV is still required.
This result suggests the existence of a very hot ($kT \sim 20$ keV) 
component around the cluster center.

The XIS spectra of the 17$'$ offset pointing can be fitted using the $1kT$ model
with $kT \sim 7.5$ keV, while that of the southern one third of 
the NWR pointing (the 1/3 NW-south region) requires a cooler value, around 4 keV.
The PIN signal in the NWR pointing, as well as the $17'$ offset pointing, can be explained
by the ICM contribution with either the $2kT$ picture suggested by the center
pointing or the $1kT$ picture fitted to the $17'$ offset XIS data.
The shape of the PIN spectra of the center and 17$'$ offset pointings 
are too soft to account for the $\Gamma = 2.1$ IC emission suggested from the radio observation.
Thus, the wide-band {\it Suzaku} X-ray spectra do not show any signature of the IC emission 
associated with the strong ($3.7$ Jy at 1.4 GHz) radio synchrotron emission.

Using the HXD, the overall upper-limit 10--40 keV flux on the non-thermal emission
within a $34'\times34'$ FOV around the north-west radio relic is derived to be
$5.3\times 10^{-12}$ erg s$^{-1}$ cm $^{-2}$ with the $1kT$ ICM modeling.
The ICM modeling using the {\it XMM}-pn spectra gives a similar upper limit. 
Using the XIS, the upper-limit on the IC emission from the radio relic itself 
is derived to be $3.9\times10^{-13}$ erg s$^{-1}$ cm $^{-2}$, %lin 3.6
when converted into the 10--40 keV band and assuming $\Gamma = 2.1$.
If we allow the index to be free, i.e. assume that there is a spectral break
somewhere above 10 keV, the upper-limit value became $7.3\times10^{-13}$ erg s$^{-1}$ cm$^{-2}$.

\subsection{The magnetic field in the north-west radio relic}
\label{chap:discussion-nwr}

By combining the radio and X-ray observations, here we estimate the magnetic field strength
in the north-west radio relic.
From the radio observation, the relic has
a flux of 3.7 Jy at 1.4 GHz (Johnston-Hollit 2008).
Its radio fluxes at 85 MHz, 843 MHz, 1.4 GHz and 2.3 GHz
are reported to roughly follow the power law with $\Gamma = 2.1$ (Roettgering et al. 1997).
With a magnetic field of $1~\mu$G, the 85 MHz radio emission
requires electrons with a Lorentz factor $\gamma$ of $\sim 5\times10^3$,
which in turn emits IC photon at around 20 keV.
In the case of 1.4 GHz radio emission, they are $\gamma \sim 2\times10^4$ and 350 keV, respectively.
Thus, we will assume that the photon index is similar to $\Gamma = 2.1$ above 20 keV,
when the magnetic field is stronger than 1 $\mu$G.

The radio flux at the 80--320 MHz band,
which corresponds to the $\sim 20$ keV IC emission with $B = 1~\mu$G,
is derived as $F^{\rm Sync}_{\rm 80-320~MHz} = 1.8\times10^{-13}~{\rm erg~ s^{-1}~ cm^{-2}}$,
% 1.5 in 0.4-1.6 GHz
assuming a $\Gamma =2.1$ power-law emission and a flux of 3.7 Jy at 1.4 GHz.
The XIS upper-limit right on the north-west radio relic region is
derived as $F^{\rm IC}_{\rm 10-40~keV} < 3.9\times 10^{-13}~{\rm erg~ s^{-1}~ cm^{-2}}$,
again assuming a simple $\Gamma =2.1$ power law.
By dividing these two values, we obtain 
$$\frac{F^{\rm Sync}}{F^{\rm IC}} = \frac{U_{\rm B}}{U_{\rm CMB}}$$
$$= (\frac{B}{B_{eq.}^{\rm CMB}})^2
 > \frac{1.8\times10^{-13}}{3.9\times10^{-13}} .$$ %lin: 3.5
Here, $B_{eq.}^{\rm CMB} = 3.2~\mu$G is the equivalent magnetic field of the energy density of the CMB.
Thus, the magnetic field in the north-west radio relic is inferred to be $B > 2.2~\mu$G. %lin OK

Since the radio emission is detected only down to 85 MHz,
the IC spectra below $\sim 20$ keV might have a break.
For the electrons with $\gamma = 5\times10^3$ within 
a thermal gas density of $\sim 1.3\times10^{-4}$ cm$^{-3}$ (derived from the XIS results),
the dominant cooling processes are the IC and the synchrotron emissions.
These give a cooling time of $\sim 4 \times 10^8$ years.
If the duration of electron injection is shorter than this time-scale,
the electron distribution below this energy is not affected by the cooling,
and the resultant IC (and synchrotron) photon index can be harder, typically $1.5 < \Gamma < 2.1$.
From the $\Gamma$ free power-law fitting to the XIS spectra obtained from the north-west relic region,
we know that the upper-limit IC flux in the 10--40 keV band
is $7.3\times 10^{-13}~{\rm erg~ s^{-1}~ cm^{-2}}$.
In this model, the photon index is 1.6 around 4 keV, and breaks to 2.1 above $\sim 20$ keV. 
In this case, the magnetic field strength is inferred to be $B > 1.6~\mu$G. 
These results are consistent with the radio rotation measure observations
of this relic, suggesting 3--5 $\mu$G (Johnston-Hollitt 2004).

Based on the observable parameters, we calculate the energy density of
individual components in the north-west radio relic.
The thermal particle density and temperature give 
an energy density $U_{\rm th}$ of 1.0 eV cm$^{-3}$. 
Here, we simply assume that all X-rays in the north-west relic are of thermal origin.
With a magnetic field strength of $> 2.2$ $\mu$G and $>1.6$ $\mu$G,  %lin 1.5
the energy density $U_{B}$ is derived as $> 0.12$ eV cm$^{-3}$ or $>0.07$ eV cm$^{-3}$, %lin 0.06
respectively.
Here, on the contrary, we assume that most of the X-rays are of IC origin
to give an upper limit on $U_e$.
If a power-law distribution of relativistic electrons are assumed
within $5\times10^2 < \gamma  < 4\times10^4$, the energy density of electrons $U_e$ 
is $< 0.15$ eV cm$^{-3}$. Here the minimum $\gamma$
is assumed to be the critical Lorentz factor below where ionization losses dominate (Sarazin 1999).
Although $U_e$ strongly depends on the 
unknown electron distribution below $\gamma \sim 5\times10^3$,
the calculated upper-limit value is not so far from the equi-partition condition.

The ratio between the non-thermal 
($U_{B}$ and $U_{e}$) and thermal ($U_{\rm th}$) energy density  is inferred to be as large as 20\%.
Although $U_e$ has a large uncertainty, 
the lower limit on $U_B$ is much more robust.
In addition, $U_{B}/U_{\rm th} > 0.07$ is a strong limit. 
If we assume, for example, that about a half of the X-rays in the north-west 
relic are of thermal origin and
the latter half is of IC origin, 
$U_{\rm th}$ is decreased by a factor of $\sqrt{2}$ and the lower limit on 
the $U_{\rm B}$ is doubled. In this case, the ratio becomes as high as 19\%,
excluding the $U_e$ contribution.
Our results are generally consistent with those by Rottgering et al. (1997),
though they do not discuss in detail the magnetic field.

This demonstrates that magnetic fields stronger than 1 $\mu$G exist
on scales as large as the cluster virial radius and 
exert non-negligible pressure compared to the thermal one,
at least around the region of the north-west relic.
Recent observations of radio relics around other clusters have also revealed
similarly strong magnetic fields of $> 0.8~\mu$G 
(e.g. Henriksen \& Mushotzky 2001; Feretti \& Neumann 2006; Hubert et al. 2008).
That of the Abell 3667 north-west relic is the strongest yet observed.

It is important to note that the large magnetic field we have
found applies only to the interior of the radio relic.  Thus,
it is possible that most clusters do not have such large fields
at such outer radii.  
The relic is the brightest diffuse
cluster radio source known, and its interior may not be typical
of the diffuse intracluster medium outside the relic, or in other clusters.  
It may be that the magnetic field and seed relativistic particles
were injected into this region by an AGN in the past
(e.g. Ensslin \& Gopal-Krishna 2001).
If the relativistic electrons were accelerated by
a merger shock, the same shock may have amplified the magnetic
field, perhaps by some mechanism like the Weibel instability 
(e.g. Okabe \& Hattori 2003; Fujita et al.~2006; Kato 2007)
or the cosmic ray streaming instability (Lucek \& Bell 2000, Bell 2004).
Also, Abell~3667 has undergone a very dramatic merger,
and may not be typical of more relaxed clusters.

\subsection{Comparison of the HXD-PIN upper-limit of the IC emission with other results}
\label{chap:discussion-whole}

The HXD-PIN data of the NWR pointing provides an upper-limit
on the IC emission as $5.3\times10^{-12}$ erg s$^{-1}$ cm$^{-2}$.
Since the FOV of the PIN is $\sim 10$ times larger than the region used in the XIS analysis,
it provides valuable limits on any non-thremal electrons that may be 
distributed on scales larger than the north-west relic.
These electrons, in addition, are not readily observable in the radio band.
For example, if the relic is surrounded by a larger region with lower
magnetic field, say $0.3~\mu$G, electrons therein with 
$\gamma \sim 5\times10^3$ would be unobservable through their 
synchrotron emission since its frequency is less than 10 MHz.

Using the {\it Beppo-SAX} PDS data with an equivalent exposure (113 ks) to the NWR pointing,
Fusco-Femiano et al.~(2001) derived the upper-limit hard X-ray flux of
$\sim 6.4\times 10^{-12}$ erg s$^{-1}$ cm$^{-2}$ in the 20--80 keV band
assuming a $\Gamma =2.1$ power-law emission.
Here, they fixed the ICM temperature to 7 keV and included only the statistical errors.
Using the same data with similar thermal model,
Nevalainen et al.~(2004) presented a 90\% confidence upper limit flux of
$7.1\times10^{-12}$ erg s$^{-1}$ cm$^{-2}$ in the same energy band 
assuming a power-law with $\Gamma =2.0$.
The main difference in the two results is that the latter author took into account 
the possible systematics in the PDS background modeling and 
AGN contribution derived in their own manner.

By comparing the observational results on the Crab,
we found that the 20--80 keV flux derived from {\it Beppo-SAX} PDS is
21\% smaller than those derived from {\it Suzaku}.
Thus, results from Fusco-Femiano et al.~(2001) and Nevalainen et al.~(2004) 
can be converted into {\it Suzaku}-equivalent fluxes 
of $8.1$ and $9.0\times10^{-12}$ erg s$^{-1}$ cm$^{-2}$, respectively.
Therefore, our {\it Suzaku} result provides more stringent upper limit on the flux,
compared to these two results.
Furthermore, it is focused on a more spatially restricted region around the north-west radio relic,
thanks to the narrow FOV of the HXD ($\phi \sim 34'$ FWHM)
compared to that of the PDS ($\phi \sim 1.3$\degree~FWHM).

\subsection{Possible very hot component in the merging cluster}
\label{chap:discussion-hot}

The HXD-PIN results from the cluster center shows a ``soft-shaped'' hard component
with $\Gamma = 3.6_{-0.9}^{+1.1}$, which can be best interpreted as 
a very hot component with a temperature of $19.2_{-6.0}^{+6.6}$ keV, 
in addition to the $\sim 4.7$ keV component in the $2kT$ picture combined with the XIS spectra.
Using the $3kT$ model, 
the temperature of the very hot component is required to be higher than 20 keV.
Since the spectra can be marginally fitted with a $\Gamma \sim 1.4$ power-law component as well,
the emission could be of non-thermal origin, 
or contamination from a yet unknown Compton-thick AGN.
However, our contaminating point-source survey (appendix \ref{apdx:point_src})
and the softness of the spectral shape itself suggests the thermal interpretation.
Since the excess signal to the $1kT$ component is significant in the center pointing,
the very hot component can be significant around the cluster center.
Detailed modeling of the ICM using all the available data,
including our new XMM data to be obtained soon, will help clarifying the actual value.

From the {\it ASCA} observations, 
this cluster is known to host significant temperature inhomogeniety,
showing regions with a temperature as high as 10 keV 
together with the 4--6 keV dominant emission (Markevitch 1998).
Later, Briel, Finoguenov \& Henry (2004) also showed the complex 
multi temperature nature of the ICM using the {\it XMM} data. 
The temperature ranges from 4 keV to 8 keV,
and several hot regions are visible around the cluster center.
Since these analyses are based on the projected X-ray color maps,
the temperature of the actual hottest component can be higher 
considering the convolution along the line-of-sight.
The {\it Beppo-SAX} PDS spectrum is consistent with the center-$2kT$ model
taking into account all the systematic errors. 
Thus, our results on ``soft-shaped'' hard component is basically consistent with existing data.

Such a very hot component, with a temperature as high as 20 keV,
is suggested by many hydrodynamical simulations of merging clusters 
right after the time of impact (e.g. Takizawa 2000; Ricker \& Sarazin 2001).
Since the Abell~3667 cluster is considered to be at this stage,
the very hot component will be the hottest heated ICM emission.
We note that a similar very hot component is suggested in 
the cluster RXJ1347.5--1145 (Ota et al. 2008) using the {\it Suzaku} wide-band data combined 
with {\it Chandra} spectra.
Its high temperature and relatively large luminosity means that
the very hot component cannot be in pressure balance with the cooler one,
otherwise, it will dominate the volume in the cluster.
Thus, the emission should be temporary 
originating from the shock and/or the compression heating,
and is expected to disappear soon by adiabatic expansion.
To investigate the location of the emission, we will need 
much wider energy band coverage in the imaging detector,
specifically, arcmin-scale imaging spectroscopy up to $\sim 40$ keV.

\section{Conclusion}
\label{chap:conclusion}

From {\it Suzaku} mapping observations of Abell~3667, upper-limits on 
the non-thermal (hard) X-ray emission were obtained.
On the north-west radio relic itself, the XIS limit 
of 7.3$\times10^{-13}$ erg s$^{-1}$ cm$^{-2}$ extrapolated to the 10--40 keV band
provides a lower-limit magnetic field of $> 1.6~\mu$G.  %1.5
The non-thermal energy density there is estimated to be higher than 7\%, and likely near 20\%.

By combining the XIS and the {\it XMM} mapping data to the HXD-PIN spectra,
the upper limit on the 10--40 keV excess hard emission is derived as 
%3.2--7.0
1.0--5.3$\times10^{-12}$ erg s$^{-1}$ cm$^{-2}$, depending on the modeling of the ICM.
Since the north-west relic has a magnetic field of $> 1$ $\mu$G
and hence must be almost an order of magnitude darker than this hard X-ray limit,
the relativistic electrons must be distributed in regions other than the relic
with low magnetic field (of $B < 0.3~\mu$G) if IC hard X-ray emission near the flux limit exists.

The PIN data from the pointing on the cluster center shows hard X-rays in excess to 
the averaged ICM emission, which is thought to have a temperature of around 7~keV.
Although other interpretations cannot be rejected,
its soft spectral shape indicates the emission to be possibly from a very hot thermal component
with a temperature of $\sim 20$ keV, generated temporary in the cluster merger phase.

In the near future, X-ray observatories with hard X-ray imaging optics operating up to 40 keV
will be launched, such as the ASTRO-H (former NeXT; e.g. Takahashi et al 2006),
NuSTAR and Simbol-X (e.g. Ferrando et al. 2006).
These detectors will drastically improve the sensitivity
to both point sources and relatively clumpy extended hard X-rays,
such as the local non-thermal emission associated with
the radio halos/relics and the very hot component.
In addition, the Soft Gamma-ray Detector onboard ASTRO-H,
(a collimated instrument) features the highest sensitivity at around 100 keV
using the novel narrow-field-of-view Compton camera concept.
It will open a new window to the widely distributed non-thermal emission, which is difficult to separate from 
the ICM thermal emission below $\sim 50$~keV.
\\

K.N. and K.M. are supported in part by a Grant-in-Aid from the Ministry of
Education, Science, Sports, and Culture of Japan (18104004),
and M.T. (16740105,19740096) and S.I (19047004, 19540283), as well.
C.L.S. and D.R.W. were supported by NASA {\it Suzaku} grants
NNX06AI44G and NNX06AI37G and {\it XMM-Newton} grant NNX06AE76G.
D.R.W. was also supported by a Dupont Fellowship and a Virginia
Space Grant Consortium Fellowship.  A.F. was partially supported
by NASA grant NNG05GM5OG to UMBC.
Basic research in radio astronomy of T.C. at the NRL is supported by 6.1 Base funding.

\appendix

\section{The CXB estimation for the XIS}
\label{apdx:xis_cxb}

The Lockman Hole observation ({\it Suzaku} observation ID, 101002010) 
was used as the CXB template.
The NXB subtracted data from the Lockman Hole were fitted with a model
composed of a power-law with a fixed $\Gamma$ of 1.4, 
and thermal components (so called ``Galactic components'') with
one of the temperatures fixed at 0.08 keV and another fitted to give $\sim 0.3$ keV.
The 2--10 keV flux of the power-law component was derived as
$5.7\pm0.1 \times 10^{-8}$ erg s$^{-1}$ cm$^{-2}$ str$^{-1}$,
which is consistent within 1 $\sigma$ from the best fit value derived using {\it ASCA} (Kushino et al. 2002).

We also analyzed the XIS data of the outer boundary for Abell 3667,
using the northern 1/3 region of the NWR pointing (hereafter the 1/3 NW-north region).
It gave a $\sim$ 16\% higher normalization for the power-law component,
which is consistent with the expected fluctuation of CXB, discussed later in this section.
The ``Galactic component'' parameters are similar,
but the flux is brighter by a factor of 2--5,
which is also consistent with its sky fluctuations.
Because the 1/3 NW-north region could be contaminated by the 
outskirts of the cluster ICM emission,
we modeled the CXB with the power-law component fixed at the value obtained from the Lockman Hole observation, while the ``Galactic component''  was fixed at that of the 1/3 NW-north region.

The CXB fluctuations can be modeled as
$\sigma_{\rm CXB} / I_{\rm CXB} \propto \Omega_e^{-0.5} S_c^{0.25}$.
Here, $\Omega_e$ is the effective solid angle and $S_c$ is the upper cutoff flux.
From the HEAO-1 A2 results, Shafer (1983) derived
$\sigma_{\rm CXB} / I_{\rm CXB} = 2.8$\%, with 
$\Omega_e = 15.8$ deg$^2$ and  $S_c = 8\times 10^{-11}$ erg s$^{-1}$ cm$^{-2}$.
By scaling this result with the XIS parameters,
i.e. $\Omega_e = 0.09$ deg$^2$ and  $S_c \sim 1\times 10^{-13}$ erg s$^{-1}$ cm$^{-2}$
in the cluster vicinity, we obtain the CXB fluctuation over the XIS full FOV as 11\% (90\% confidence level).
For the NWR 1/3 NW-north region, the value is then 19\%.
For the north-west relic region with $\Omega_e = 0.04$ deg$^2$
and $S_c \sim 2\times 10^{-14}$ erg s$^{-1}$ cm$^{-2}$, it becomes 15.9\%.
Note that this scaling relation is consistent with the {\it ASCA} and {\it Ginga} CXB analysis (Ishisaki 1997).

\section{The CXB estimation for the HXD}
\label{apdx:hxd_cxb}

The CXB flux in the PIN is modeled as a cutoff power-law, described as
$N(E) = n \times (E)^{-\Gamma} \times {\rm exp}(\it -E/E_{\rm fold})$ 
in ${\rm photons~ cm^{-2}~s^{-1}~keV^{-1}~FOV^{-1} }$.
Here, $E_{\rm fold}$ is an folding energy, and $n$ is the normalization.
In energies below $\sim 80$~keV, the HXD energy response
for a diffuse source is the same to that for a point source except for its normalization.
Thus, we simply scaled our CXB model to the solid angle of the HXD, 
$\Omega_e^{\rm HXD} = 0.32$ deg$^2$ and utilized the HXD nominal response.
We first used the parameters based on the HEAO-1 results (Boldt 1987),
which gave $\Gamma = 1.29$, $E_{\rm fold} = 40.0$ and $n= 8.36\times10^{-4}$. 
We then compared this model to the XIS observation of Lockman Hole,
on which our XIS CXB modeling is based on.
It was found that the model predicts a flux 8.6\% lower in the XIS 3-8 keV band data.
Thus, we modified the PIN CXB model normalization to match the difference, as $n= 8.69\times10^{-6}$. 
Note that although there is an apparent difference in photon index
between the CXB model of the XIS and PIN, the cutoff function in the later makes 
two models perfectly matched in the energy band at 3--8 keV.
Through the same approach applied to the XIS data,
the CXB fluctuation in the HXD FOV is expected to be $18$\%,
with $\Omega_e^{\rm HXD} = 0.32$ deg$^2$ and 
a conservative upper cutoff flux of 
$S_c \sim 8\times 10^{-12}$ erg s$^{-1}$ cm$^{-2}$ in the 10--40 keV band.

%--------- 4.4 ------------------------
\section{Estimation of point source contamination in the FOV of the PIN}
\label{apdx:point_src}
%--------- 4.4 ------------------------

%%#################### figure 2 #########################################
\begin{figure}
  \begin{center}
    \FigureFile(7.5cm,7.5cm){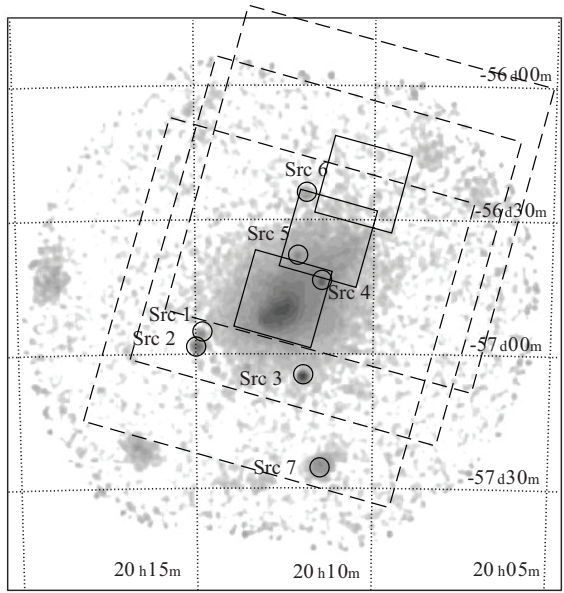} 
\end{center}
\caption{The background subtracted {\it ROSAT} PSPC image ($50'$ radius)
of Abell~3667. Gray-scale contour is logarithmic scale.
The FOVs ($68'\times68'$ bottom to bottom) of PIN are shown in dashed thin rectangles, 
those of the XIS are shown in solid rectangles,
and X-ray point sources in the FOVs are shown in circles.
}
\label{fig:pspc_src}
\end{figure}
%%#######################################################################

Since the PIN has a wider FOV ($68'\times 68'$ bottom to bottom) 
than that of the XIS ($18' \times 18'$) and also lacks imaging capability, 
signal contamination from point sources within the FOV should be carefully estimated.
Figure~\ref{fig:pspc_src} shows the PSPC image of Abell~3667
with the FOVs of the PIN superposed. 
The PSPC data is processed through the method described by Snowden et al.~(1994).
There exists seven X-ray point sources in the PIN FOVs. We identified these sources using
the SIMBAD astronomical database.% as listed in table~\ref{table:pspc_src}.

Source 4 and Source 6 are in the FOV of XIS,
and thus their flux is estimated by the XIS data.
Although the other sources are outside the FOV of XIS, 
archival data of {\it ASCA} GIS are available for the
sources 1, 2, 3, and 5. We therefore extracted the GIS spectra of source
1 thorough source 6 and fitted them with a power low modified by galactic absorption.
Blank sky observations of the GIS were used for the background of source 1,2 and 3,
while neighboring field is used for source 5.

Using the best-fit parameters, we simulated the PIN spectra
of the six sources for the three pointings,
considering the PIN transmission to each of them on the three pointings.
The total contribution of the six point sources thus estimated is more than 
an order of magnitude lower than the detected signal, and hence negligible.
Source 7, not included in this estimation, is located at the boundary 
of the PIN FOV of the center pointing. It is also a weak hard X-ray source, 
since it is a distant cluster of galaxies. 

%%#################### figure 1 #########################################
%\begin{figure}
%  \begin{center}
%    %\FigureFile(80mm,80mm){./fig/n1550_dss_xmm.ps}
%    %\FigureFile(80mm,80mm){./fig/mkimage2.ps}
%\includegraphics[width=0.33\textwidth,clip]{./fig/080711_pinall_point_1.eps}
%%\includegraphics[width=0.23\textwidth,angle=-90,clip]{./fig/src_center.ps}
%%\includegraphics[width=0.23\textwidth,angle=-90,clip]{./fig/src_17off.ps}
%%\includegraphics[width=0.23\textwidth,angle=-90,clip]{./fig/src_nwr.ps}
%       %%% \FigureFile(width,height){filename}
%  \end{center}
%  \caption{The 13--40 keV background (NXB+CXB) subtracted PIN spectra (crosses) of 
%center (top), $17'$ offset (middle), and NWR (bottom) pointings
%with estimated contributions of the point sources (solid). }
%\label{fig:point_spec}
%\end{figure}
%%%#######################################################################

%\clearpage


\begin{thebibliography}{}

\bibitem[anders(1989)]{key-5}
Anders E., \& Grevesse N. \ 1989, \gca, 53, 197 

\bibitem[Bell et al.(2004)]{Bell04}
Bell A. R.\ 2004 \mnras 353, 550

\bibitem[Briel et al.(2004)]{BFH04}
Briel, U. G., Finoguenov, A., Henry, J. P., \&  Anders E., \ 2004, \aap, 426, 1

\bibitem[Boldt(1987)]{key-5}
Boldt E. \ 1987, \iaucirc 124, 611

\bibitem[carilli(2002)]{key-5}
Carilli C. L. \& Taylor G. B., \ 2002 \araa, 40, 319

\bibitem[Clarke et al. (2001)]{key-5}
Clarke T. E., Kronberg P. P. \& B\"ohringer H. \ 2001, \apjl, 547, 111

\bibitem[Dickey \& Lockman 1990]{key-5}
Dickey  J. M. \& Lockman F. J. \  1990, \araa, 28, 215 

\bibitem[Ensslin \& Gopal-Krishna (2001)]{key-5} %Coma relic
Ensslin, T. A. \& Gopal-Krishna \ 2001\aap, 366, 26

\bibitem[Feretti \& Neumann (2006)]{key-5} %Coma relic
Feretti L., Neumann D.M., \ 2006 \aap, 450, L21

\bibitem[Ferrando Simbol-X(2006)]{key-5} %Simbol-X
Ferrando P. et al. \ 2006, SPIE proc., 6266

\bibitem[Fujita \& Kato (2006)]{key-5}
Fujita Y., Kato T. N. \& Okabe N. \ 2006, Phys. Plasma 13, 122901

\bibitem[fukazawa_h62(2000)]{key-5}
Fukazawa Y., Nakazawa K., Isobe N., Makishima K., Matsushita K.,
Ohashi T. \& Kamae T. \ 2001\apjl, 546, 87

\bibitem[fukazawa_asca(2004)]{key-5}
Fukazawa Y., Makishima K., \& Ohashi T. \ 2004, \pasj, 56, 965

\bibitem[fukazawa_asca(2008)]{key-5}
Fukazawa Y. et al. \ 2008, submitted to \pasj

\bibitem[ff_coma(1999)]{key-5}
Fusco-Femiano, R., dal Fiume D., Feretti L., Giovannini G., Grandi P.,
Matt G., Molendi S. \& Santangelo A. \ 1999, \apjl, 513, 21

\bibitem[ff_a3667(2001)]{key-5}
Fusco-Femiano R., Dal Fiume D., Orlandini M., Brunetti G., Feretti L., \& Giovannini G. \ 2001, \apj, 552, L97

\bibitem[ff_coma(2007)]{key-5}
Fusco-Femiano R., Landi R. \& Orlandini M. \ 2007, \apj, 654, L9

\bibitem[Henriksen et al. (2001)]{key-5} % A1367
Henriksen M. \& Mushotzky R. \ 2001, \apj, 553, 84

\bibitem[Hubert et al. (2008)]{key-5}
Hubert C. M., Harris D. E., Harrison F. A. \& Mao P. H. \ 2008, \mnras, 383, 1259

\bibitem[Ishisaki (1997)]{key-6}
Inoue S., Aharonian F. A. \& Sugiyama N.\ 2005, \apj, 628, L9

\bibitem[Ishisaki (1997)]{key-6}
Ishisaki, Y. 1997,  Ph.D. thesis, University of Tokyo

\bibitem[johnston-hollitt(2004)]{key-5}
Johnston-Hollitt M. \ 2004, in Proc. The Riddle of Cooling Flows in Galaxies and Clusters of Galaxies, ed. T. H. Reiprich, J.C. Kempner, \& N. Soker, 51

\bibitem[johnston-hollitt(2008)]{key-5}
Johnston-Hollitt M., Hunstead R. W. \& Corbett, E. \ 2008, \aap, 479, 1

\bibitem[Kato (2007)]{key-5}
Kato T. N. \ 2007, \apj, 668, 974

\bibitem[Kitaguchi_cluster (2008)]{key-5}
Kitaguchi T. et al. \ 2008, submitted to \pasj

\bibitem[knopp_pspc_a3667(1996)]{key-5}
Knopp G. P., Henry J. P., \& Briel U. G., \ 1996, \apj, 472, 125

\bibitem[kokubun_hxd(2007)]{key-5}
Kokubun M., et al. \ 2007, \pasj, 59, 53

\bibitem[koyama_xis(2007)]{key-5}
Koyama K., et al. \ 2007, \pasj, 59, 23

\bibitem[kushino_cxb(2007)]{key-5}
Kushino A., Ishisaki Y., Morita U., Yamasaki N. Y., Ishida M., Ohashi T. \& Ueda Y. \ 2002, \pasj, 54, 327

\bibitem[Lucek \& Bell (2000)]{LB00}
Lucek S. G. \& Bell A. R.\ 2000 \mnras 314, 65

\bibitem[maxim_a3667(1999)]{key-5}
Markevitch M., Sarazin C. L., \& Vikhlinin A. \ 1999, \apj, 521, 526
%Physics of the Merging Clusters Cygnus A, A3667, and A2065

\bibitem[mazzotta_a3667(1998)]{key-5}
Mazzotta P., Fusco-Femiano R., \& Vikhlinin A. \ 2002, \apj, 569, L31
%Chandra Observation of a 300 Kiloparsec Hydrodynamic Instability in the Intergalactic Medium of the Merging Cluster of Galaxies A3667

\bibitem[mitsuda_suzaku(2007)]{key-5}
Mitsuda K., et al. \ 2007, \pasj, 59, 1

\bibitem[mohr et al (1999)]{key-5}
Mohr J. J., Mathiesen B. \& Evrard A. E.\ 1999, \apj, 517, 627

\bibitem[nakazawa_group(2006)]{key-5}
Nakazawa K., Makishima K., \& Fukazawa Y. \ 2007, \pasj, 59, 167

\bibitem[nevalinen_pds(2004)]{key-5}
Nevalainen J., Oosterbroek T., Bonamente M., \& Colafrancesco S., \ 2004, \apj, 608, 166

\bibitem[neemann_arnaud_1999]{key-5}
Neumann D. M. \& Arnaud M. \ 1999, \aap, 348, 711

%\bibitem[markevitch_a3667(1998)]{key-5}
%Markevitch M., \ 2004, AN, 320, 185

%\bibitem[repha_a3667_xte(2004)]{key-5}
%Rephaeli Y. \& Gruber D., \ 2004, \apj, 606, 825
%%Spectral Analysis of Rossi X-Ray Timing Explorer Observations of A3667

\bibitem[okabe_hattori(2003)]{key-5}
Okabe N. \& Hattori M. \ 2003, \apj, 599, 964

\bibitem[ota_et al. 2008]{key-5}
Ota N., Murase K., Kitayama T., Komatsu K., Hattori M., Matsuo H., Oshima T.,
Suto Y. \& Yoshikawa K. \ 2008, arXiv:0805.0500

\bibitem[Petrosian et al. 2006]{key-5}
Petrosian V., Madejski G. \& Luli K. \ 2006, \apj, 652, 948

\bibitem[Read \& Ponman (2003)]{RP03}
Read A. M. \& Ponman T. J. \ 2003, \aap, 409, 395

\bibitem[Rephaeli et al. 2006]{repha03}
Rephaeli Y., Gruber D., \& Arieli Y. \ 2006, \apj, 649, 673

\bibitem[Ricker \& Sarazin 2001]{RS01}
Ricker P. M. \& Sarazin C. L.,\ 2001, \apj, 561, 621

\bibitem[roettiger_origin(1999)]{key-5}
Roettiger K., Burns J. O., Stone J. M., \  1999, \apj, 518, 603
%A Cluster Merger and the Origin of the Extended Radio Emission in Abell 3667

\bibitem[Rosetti_and_Molendi_2004]{key-5}
Rossetti, M. \& Molendi, S. \ 2004, \aap, 414, L41

\bibitem[roettgering_radio(1997)]{key-5}
Rottgering H. J. A., Wieringa M. H., Hunstead R.W., \& Ekers R. D., \ 1997, \mnras, 290, 577
%The extended radio emission in the luminous X-ray cluster A3667

\bibitem[sarazin(1988)]{key-5}	
Sarazin C. L. \ 1988, X-ray emission from clusters of galaxies, (Cambridge University Press)

\bibitem[Sarazin (1999)]{key-5}	
Sarazin C. L. \ 1999, \apj, 520, 529

\bibitem[serlemitsos_xrt(2007)]{key-5}
Serlemitsos P.J., et al. \ 2007, \pasj, 59, 9

\bibitem[Shafer (1983)]{key-5}	
Shafer, R. A. \ 1983, PhD. Thesis, University of Maryland

\bibitem[Snowden_pspc(1994)]{key-5}
Snowden, S. L., McCammon, D., Burrows, D. N., \& Mendenhall, J. A. \ 1994, \apj, 424, 714

\bibitem[sodre_a3667opt(1992)]{key-5}
Sodre L. Jr., Capelato H. V., Steiner J. E., Proust D., \& Mazure A.  \ 1992, \mnras, 259, 233

\bibitem[struble \& Herbert (1999)]{key-5}
Struble M. F. \& Rood H. J. \ 1999, \apj, 125, 35

\bibitem[takahashi_hxd(2007)]{key-5}
Takahashi T., et al. \ 2007, \pasj, 59, 35

\bibitem[takahashi_next(2006)]{key-5}
Takahashi T., Mitsuda K. \& Kunieda H. \ 2006, SPIE proc., 6266 


\bibitem[takizawa_sph(2000)]{key-5}
Takizawa M. \ 2000, \apj, 532, 183

\bibitem[tawa_nxb(2007)]{key-5}
Tawa N. et al. \ 2008, \pasj, 60, 11

\bibitem[vikhlinin_coldfront2(2001)]{key-5}
Vikhlinin A., Markevitch M., \& Murray S. S. \ 2001, \apj, 551, 160

\bibitem[vikhlinin_coldfront1(2001)]{key-5}
Vikhlinin A., Markevitch M., \& Murray S. S. \ 2001, \apj, 549, L47

\bibitem[watanabe_coma(1999)]{key-5}
Watanabe M., Yamashita K., Furuzawa A., Kunieda H., Tawara Y. \& Honda H.
\ 1999, \apj, 527, 80

\bibitem[Wik_Coma(2008)]{key-5}
Wik D. R., Sarazin C. L., Finoguenov A., Matsushita K., Nakazawa K. \&
Clarke T. E. \ submitted to \apj

\end{thebibliography}
\end{document}